\documentclass[10pt,twocolumn,letterpaper]{article}

\usepackage{cvpr}
\usepackage{times}
\usepackage{epsfig}
\usepackage{graphicx}
\usepackage{amsmath}
\usepackage{amssymb}
\usepackage{makecell}
\usepackage{multirow}
\usepackage[ruled,vlined,linesnumbered]{algorithm2e}
\usepackage{booktabs}

\usepackage{mathtools}
\usepackage{amsmath}
\usepackage{amssymb}
\usepackage{amsfonts}
\usepackage{amsopn}
\usepackage{graphicx} 
\usepackage{textcomp}
\usepackage{xfrac}
\usepackage{bbm}
\usepackage{cite}
\usepackage{overpic}
\usepackage{subfig}
\usepackage{wrapfig}
\usepackage[ruled,vlined,linesnumbered]{algorithm2e}
\usepackage{multirow}

\usepackage{color}
\definecolor{green}{rgb}{0, 0.5, 0}
\definecolor{orange}{rgb}{0.8, 0.6, 0.2}
\definecolor{red}{rgb}{1.0, 0.0, 0.0}
\definecolor{teal}{rgb}{0.0, 0.4, 0.4}
\definecolor{purple}{rgb}{0.65,0,0.65}
\definecolor{saffron}{rgb}{0.95,0.75,0.2}
\definecolor{turquoise}{rgb}{0.0,0.5,0.5}
\definecolor{brown}{rgb}{0.5, 0.16, 0.16}

\usepackage{overpic}
\usepackage{currfile} 

\newcommand{\myfigurename}{\put(-3,0){\vertical{\todo{\currfiledir}}}}
\renewcommand{\myfigurename}{}
\newlength\savedwidth
\newcommand\whline[1]{\noalign{\global\savedwidth\arrayrulewidth
		\global\arrayrulewidth #1} %
	\hline
	\noalign{\global\arrayrulewidth\savedwidth}}

\definecolor{lightgray}{rgb}{0.6, 0.6, 0.6}

\newcommand{\Sec}[1]{Section~\ref{sec:#1}}

\usepackage[normalem]{ulem}

\newcommand{\textapprx}{\raisebox{0.1ex}{\texttildelow}}

\renewcommand{\paragraph}[1]{\textbf{#1.}}

\newcommand{\hidecomment}[1]{}

\DeclareMathOperator*{\argmin}{arg\,min}

\newcommand{\ps}{\mathcal{P}}
\newcommand{\nb}{\mathcal{N}}
\newcommand{\gt}{\mathcal{T}_\text{G}}
\newcommand{\gtx}{\mathcal{T}_\text{G}^x}
\newcommand{\gty}{\mathcal{T}_\text{G}^y}
\newcommand{\gtz}{\mathcal{T}_\text{G}^z}
\newcommand{\lt}{\mathcal{T}_\text{L}}
\newcommand{\xmin}{x_\text{min}}
\newcommand{\xmax}{x_\text{max}}
\newcommand{\ymin}{y_\text{min}}
\newcommand{\ymax}{y_\text{max}}
\newcommand{\zmin}{z_\text{min}}
\newcommand{\zmax}{z_\text{max}}

\usepackage{xspace}

\usepackage[pagebackref=true,breaklinks=true,letterpaper=true,colorlinks,bookmarks=false]{hyperref}

\cvprfinalcopy 

\def\cvprPaperID{6602} 

\ifcvprfinal\fi
\begin{document}

\title{Fusion-Aware Point Convolution for Online Semantic 3D Scene Segmentation}

\author{
Jiazhao Zhang$^{1,\ast}$ \quad\quad Chenyang Zhu$^{1,}$\thanks{Joint first authors}
\quad\quad Lintao Zheng$^{1}$ \quad\quad Kai Xu$^{1}$\thanks{Corresponding author: kevin.kai.xu@gmail.com}\\
$^1$National University of Defense Technology\\
}

\maketitle


\begin{abstract}\vspace{-12pt}
Online semantic 3D segmentation in company with real-time RGB-D reconstruction poses special challenges such as
how to perform 3D convolution directly over the progressively fused 3D geometric data,
and how to smartly fuse information from frame to frame.
We propose a novel fusion-aware 3D point convolution which operates directly on the geometric surface
being reconstructed and exploits effectively the inter-frame correlation for high quality 3D feature learning.
This is enabled by a dedicated dynamic data structure which organizes the online acquired
point cloud with \emph{global-local trees}.
Globally, we compile the online reconstructed 3D points into an incrementally growing coordinate interval tree,
enabling fast point insertion and neighborhood query.
Locally, we maintain the neighborhood information for each point using an octree whose construction benefits from the fast query of the global tree.
Both levels of trees update dynamically and help the 3D convolution effectively exploits the temporal coherence for effective information fusion across RGB-D frames.
Through evaluation on public benchmark datasets, we show that our method achieves the state-of-the-art accuracy
of semantic segmentation with online RGB-D fusion in $10$ FPS.
\end{abstract}\vspace{-18pt}


\vspace{-6pt}
\section{Introduction}
\vspace{-6pt}
Semantic segmentation of 3D scenes is an fundamental task in 3D vision.
The recent state-of-the-art methods mostly apply deep learning on either 3D geometric data solely~\cite{qi2017pointnet} or the fusion of 2D and 3D data~\cite{mccormac2017semanticfusion}.
These approaches, however, are usually offline, working with an already reconstructed 3D scene geometry~\cite{dai20183dmv,hou20183d}.
Online scene understanding associated with real-time RGB-D reconstruction~\cite{Izadi2011,Niessner2013}, on the other hand, is deemed to be more appealing due to the potential applications in robot and AR.
Technically, online analysis can also fully exploit the spatial-temporal information during RGB-D fusion.

\begin{figure}[t]
\centering
\begin{overpic}
[width=\linewidth]
{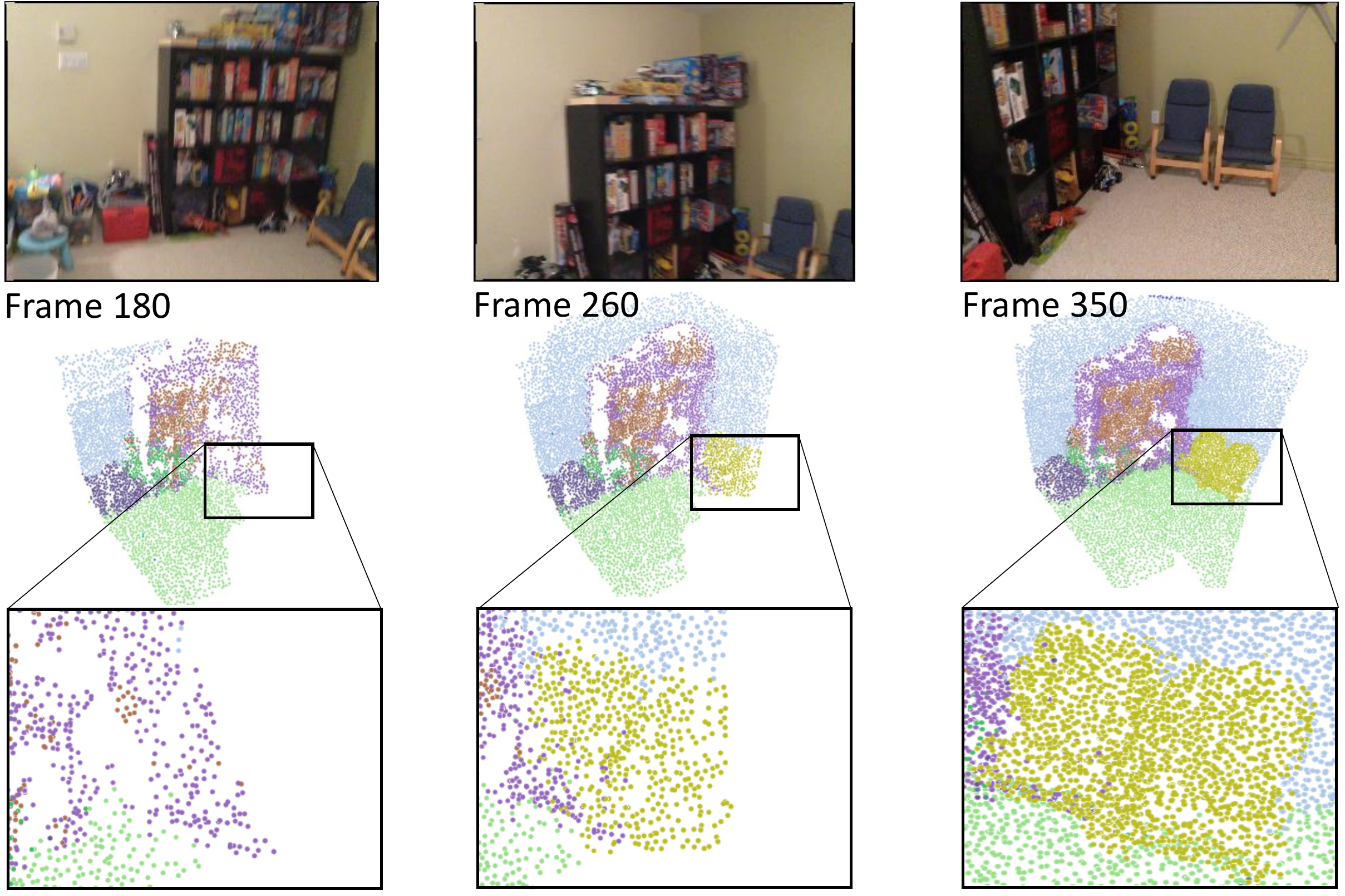}\myfigurename{}
\end{overpic}\vspace{-6pt}
\caption{
We present fusion-aware 3D point convolution which operates directly over the progressively acquired and online reconstructed scene surface. We show the point-wise labeling is being gradually improved (the chairs are recognized) as more and more frames (first row) are fused in.
}
\label{fig:teaser}\vspace{-16pt}
\end{figure}

For the task of semantic scene segmentation in company with RGB-D fusion, deep-learning-based approaches commonly adopt the \emph{frame feature fusion paradigm}. Such methods first perform 2D convolution in the individual RGB-D frames and then fuse the extracted 2D features across consecutive frames.
Previous works conduct such feature fusion through either max-pooling operation~\cite{hou20183d} or Bayesian probability updating~\cite{mccormac2017semanticfusion}.
We advocate the adoption of \emph{direct convolution over 3D surfaces} for frame feature fusion.
3D convolution on surfaces learns features of the intrinsic structure
of the geometric surfaces~\cite{bronstein2017geometric} 
that cannot be well-captured by view-based convolution and fusion.
During online RGB-D fusion, however, the scene geometry changes progressively with the incremental scanning and reconstruction. It is difficult to perform 3D convolution directly over the time-varying geometry.
Besides, to attain a powerful 3D feature learning, special designs are needed to exploit the temporal correlation between adjacent frames.

In this work, we argue that a fast and powerful 3D convolution for online segmentation necessitates
an efficient and versatile in-memory organization of dynamic 3D geometric data.
To this end, we propose a tree-based global-local dynamic data structure to enable efficient
data maintenance and 3D convolution of time-varying geometry.
Globally, we organize the online fused 3D points with an incrementally growing coordinate interval tree, which enables fast point insertion and neighborhood query.
Locally, we maintain the neighborhood information for each point using an octree whose dynamic update
benefits from the fast query of the global tree.
The local octrees facilitate efficient point-wise 3D convolution directly over the reconstructed geometric surfaces.
Both levels of trees update dynamically along with the online reconstruction.

The dynamic maintenance of the two-level trees supports 3D point convolution with feature fusion across RGB-D frames, leading to so-called \emph{fusion-aware point convolution}.
\emph{First}, point correspondence between consecutive frames can be easily retrieved from the global tree,
so that both the 2D and 3D features of a point can be efficiently aggregated from frame to frame when the point is observed by multiple frames.
\emph{Second}, with the help of per-point octrees, we realize adaptive convolution kernels at each point through weighting its neighboring points based on approximate geodesic distance. This allows a progressive
improvement of labeling accuracy across frames.

Through extensive evaluation on the public benchmark datasets,
we demonstrate that our method performs online 3D scene semantic segmentation at interactive frame-rate ($10$ FPS and even higher for key-frame-based processing) while achieving high accuracy outperforming the state-of-the-art offline methods. In particular, the accuracy achieves the top-ranking in the ScanNet benchmark,
outperforming many existing approaches including both online and offline ones.
Our main contributions include:
\begin{itemize}
\vspace{-6pt}
  \item A tree-based global-local dynamic data structure enabling efficient and powerful 3D convolution on time-varying 3D geometry.
\vspace{-6pt}
  \item A fusion-aware point convolution which exploits inter-frame correlation for quality 3D feature learning.
\vspace{-6pt}
  \item An interactive system implementing our online segmentation with real-time RGB-D reconstruction.
\end{itemize}

\begin{figure*}[!t]
\centering
\begin{overpic}
[width=\linewidth]
{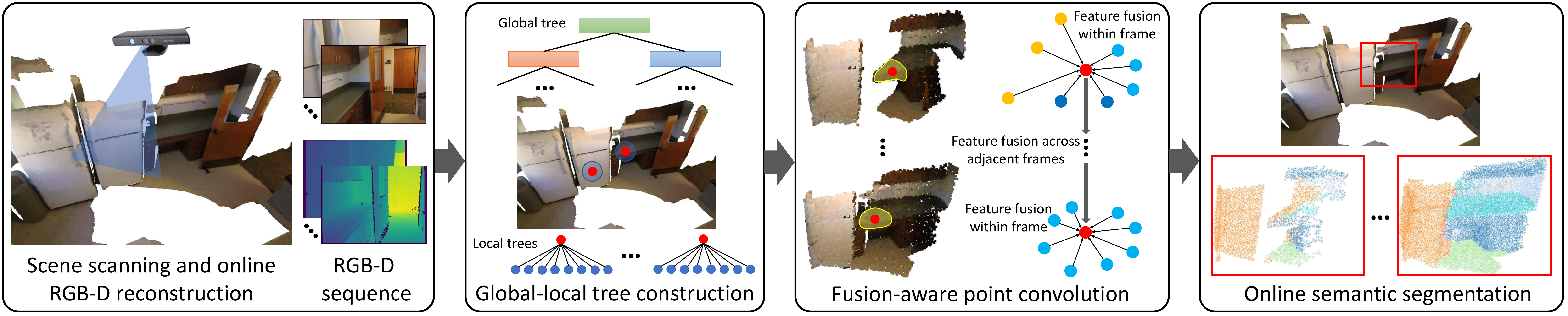}\myfigurename{}
    \put(13,-1.5){\small (a)}
    \put(38,-1.5){\small (b)}
    \put(62,-1.5){\small (c)}
    \put(87,-1.5){\small (d)}
\end{overpic}
\caption{
An overview of our pipeline. The input to our method is an online acquired RGB-D sequence being reconstructed in real-time (a). Based on the online reconstruction, we construct global-local trees to maintain a global spatial organization of the reconstructed point cloud as well as per-point local neighborhood (b). The dynamic data structure supports fusion-aware point convolution encompassing intra-frame and inter-frame feature fusion (c). Finally, the point-wise features are used for point label prediction, leading to a semantic segmentation (d).
}
\label{fig:overview}\vspace{-12pt}
\end{figure*}
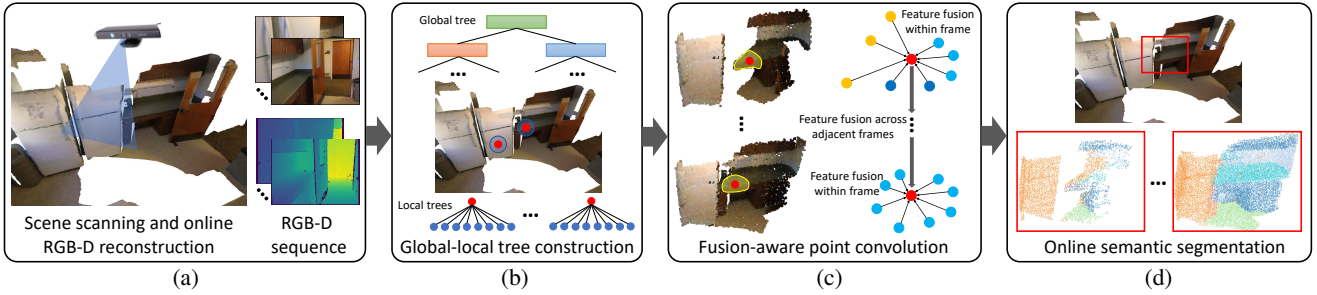


\vspace{-12pt}
\section{Related work}
\vspace{-6pt}

\paragraph{3D scene segmentation}
Scene segmentation is a long-standing problem in computer vision.
Here, we only review offline approaches handling 3D geometric data
obtained either by RGB-D fusion or LiDAR acquisition.
For fusion-based 3D reconstruction, many works~\cite{ma2017multi,lai2014unsupervised} show that the 2D labeling of the
RGB-D frames can be incorporated into the volumetric or surfel map, resulting in stable 3D labeling.
The labeling can be further improved with MRF or CRF inference over the 3D map\cite{yi2016automatic}.
These works enjoy the advances in image-based CNN for 2D segmentation.
Taking the advantage of direct 3D geometric feature learning, 3D deep learning approaches become increasingly popular,
where an efficient 3D convolution operation is the key. 
In these methods, 3D labeling is attained with CNNs operating directly on point clouds~\cite{qi2017pointnet} or their voxelization~\cite{graham20183d}.
Several other approaches conduct object detection over 3D reconstruction and then predict a segmentation mask for each detection, leading to instance segmentation~\cite{wang2018sgpn,hou20183d,yi2019gspn}.

\paragraph{Online scene segmentation}
Apart from the majority of offline batch methods,
online and incremental mapping and labeling starts to gain renewed interest lately due to the big success
of multi-view deep learning~\cite{Su15MV,qi2016,zhao2018triangle}.
Since the early attempts on 3D semantic mapping from RGB-D sequences~\cite{hermans2014dense,stuckler2014multi,Salas-Moreno2013},
a notable recent work of such kind is SemanticFusion~\cite{mccormac2017semanticfusion}.
It performs CNN-based 2D labeling for individual RGB-D frames and then probabilistically fuses the 2D predictions into a semantic 3D map. Instead of fusing prediction results, some methods~\cite{dai20183dmv,hou20183d} adopt feature map fusion based on max-pooling operation which is more deep learning friendly.
In our method, we advocate the use of 3D convolution to aggregate 2D features where a major challenge is how to handle time-varying 3D geometric data.
The DA-RNN method~\cite{xiang2017rnn} aggregates frame features using recurrent neural networks with dynamically routed connections. It smartly utilizes the data association from SLAM to connect recurrent units on the fly.
Our method, on the other hand, pursues effective direct 3D convolution through
exploiting the data association between frames.

\paragraph{Point cloud convolution}
Since 3D convolution is naturally performed on 3D Euclidean grids,
early practice opts to first converts 3D point clouds to 2D images~\cite{Su15MV} or 3D volumes~\cite{maturana2015voxnet} and then perform Euclidean convolution.
For the task of semantic segmentation, most approaches choose to
extract features in 2D and then perform segmentation in 3D based on the 2D features~\cite{ma2017multi}.
Volumetric convolution is limited by resolution due to computational cost, which
can be relieved with efficient data structure~\cite{riegler2017octnet,klokov2017escape}.
These acceleration, however, cannot handle dynamically changing point clouds like ours.
Atzmon et al.~\cite{atzmon2018point} propose a unique volume-based point convolution which consists of two
operators, extension and restriction, mapping point cloud functions to volumetric ones
and vise-versa. Point cloud convolution is defined by the extension and restriction of volumetric
convolution against the point cloud.

Since the pioneering work of PointNet~\cite{qi2017pointnet}, there have been many works focusing on
direct convolution on 3D point clouds.
Existing works aim either to improve the neighborhood structure~\cite{qi2017pointnet++,li2018pointcnn,graham20183d,su2018splatnet,tatarchenko2018tangent}
or to enhance the convolutional filters~\cite{simonovsky2017dynamic,wang2018dynamic,hermosilla2018monte,xu2018spidercnn,wu2019pointconv}.
These methods are designed to process over fixed neighborhood on static point clouds.
We design a new point convolution for time-varying geometric data with dedicated designs
targeting both aspects.
First, we maintain and update a surface-aware neighborhood structure based on a tree-based dynamic data structure. Second, we learn adaptive convolutional filters via exploiting the temporal coherence between consecutive frames.


\section{Method}

\begin{figure*}[!t]
\centering
\begin{overpic}
[width=\linewidth]
{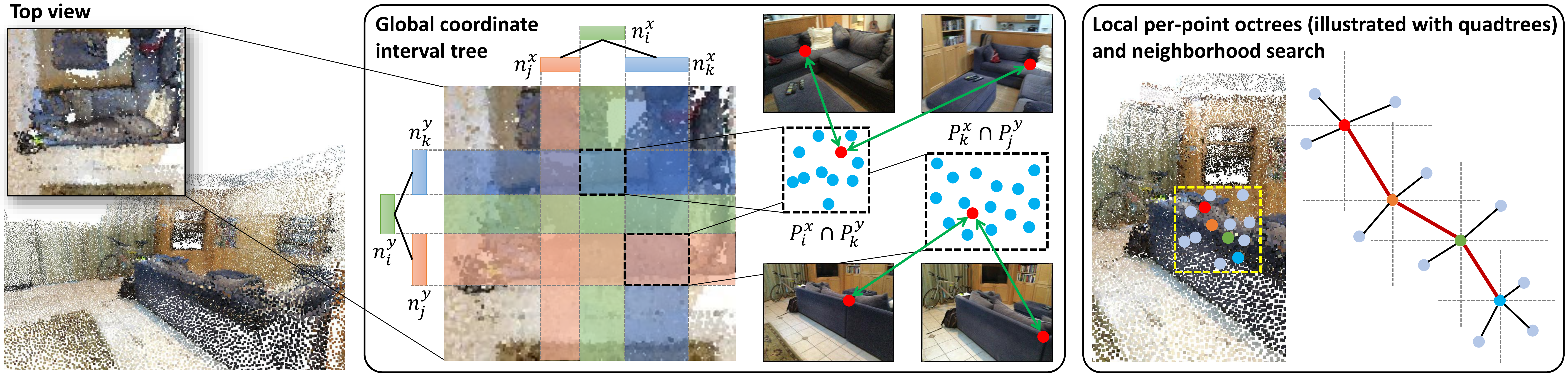}\myfigurename{}
\end{overpic}\vspace{-7pt}
\caption{
Illustration of global-local trees. The global coordinate interval trees are shown for x- and y-dimension only.
With these trees, we can find a local neighborhood for any given point as well as the correspondence between two pixels from different frames. The per-point octrees (illustrated with 2D quadtrees) can be used to find multi-ring neighborhoods.
}
\label{fig:datastructure}\vspace{-12pt}
\end{figure*}

\paragraph{Overview}
Figure~\ref{fig:overview} provides an overview of the proposed online 3D scene segmentation method.
The input to our method is an online acquired RGB-D sequence being reconstructed with real-time depth-fusion~\cite{Dai2017}.
Let us denote the RGB-D sequence by $f^k=\{(c^k_m,p^k_m)\}_{m=0}^M,k=0,1,\ldots,K$, where $c^k_m$ and $p^k_m$ store the RGB-D information of pixel $m$ of frame $k$ and the coordinates of its corresponding 3D point, respectively.
Given a reconstruction represented by a point set $\ps$, we construct a global tree $\gt$ maintaining the spatial organization of all points, as well as a per-point local trees $\{\lt(p)\}_{p\in \ps}$ storing the 1-ring neighborhood for each point (\Sec{tree}).
The dynamic data structure supports fusion-aware point convolution encompassing intra-frame and inter-frame feature fusion (\Sec{conv}). The point-wise features are used for point label prediction, resulting in a semantic segmentation (\Sec{network}).



\subsection{Dynamic Global-Local Tree Organization}
\label{sec:tree}
To support both intra-frame and inter-frame feature learning with point-based convolution, we require a data structure to organize the dynamically reconstructed, unstructured point cloud.
There are several considerations in designing such a dynamic data structure.
\emph{Firstly}, to facilitate point-based convolution, we need to construct the local neighborhood of any given point.
\emph{Second}, the data structure should support fast update of the local neighborhoods under time-varying geometry.
\emph{Thirdly}, to realize 2D-to-3D and frame-to-frame feature fusion, the data structure should allow us to find correspondence between image pixels and reconstructed points. This way, pixels across different frames can be matched through the shared corresponding 3D point.

To meet those requirements, we design a two-level tree-based data structure. Globally, we construct a coordinate-based tree organization of points which supports fast neighborhood query for any given point. This allows us to find pixel-to-point correspondence and point-based neighborhood efficiently. Based on the global tree, we build for each point an octree from which multi-scale local neighborhood can be found quickly for point-based convolution.




\paragraph{Global coordinate interval tree}
We maintain three coordinate interval trees $\gtx$, $\gty$ and $\gtz$, one for each dimension. Without loss of generality, we take $\gtx$ for example to describe the tree construction. Each node $n_i \in \gtx$ records a set of point $\ps^x_i \subset \ps$ in which each point has its x-coordinate lie in the interval $[\xmin(n_i), \xmax(n_i)]$. $\xmin(n_i)$ and $\xmax(n_i)$ are the minimum and maximum threshold for node $n_i$.
We stipulate the adjacent nodes in a coordinate interval tree complies with the following interval constraints:
%
\begin{equation*}
\xmax(n_l)<\xmin(n_p),\text{ }\xmax(n_p)<\xmin(n_r),
\end{equation*}
with $n_l$ and $n_r$ being the left and right child of node $n_p$.
The entire 3D scene is then split into slices along x-dimension.


The coordinate interval tree is constructed dynamically as more 3D points are reconstructed and inserted.
The point insertion of coordinate interval tree is conducted as follows.
Given a 3D point $p=(x_p,y_p,z_p)$, we first find a node $n_i\in \gtx$ satisfying $x_p\in[\xmin(n_i),\xmax(n_i)]$, through a top-down traverse of the tree. If such a node exists, $p$ is added to the corresponding point set $\ps^x_i$ of the node. Otherwise, we create a new leaf node whose point set is initialized as $\{p\}$ and coordinate interval as $[x_p-h, x_p+h]$. Here, $h$ is the half size of coordinate intervals.
This new node is then attached to the node whose interval is closest to the new node's interval. The detailed explanation of tree construction with balance maintenance can be found in the supplemental material.


After constructing the coordinate interval trees for all three dimensions, we can achieve efficient point correspondence search and neighborhood retrieval for any given query 3D point $q=(x_q,y_q,z_q)$.
Through traversing the three trees, we obtain three nodes $n_i \in \gtx$, $n_j \in \gty$, $n_k \in \gtz$ satisfying $x_p\in[\xmin(n_i),\xmax(n_i)]$, $y_p\in[\ymin(n_j),\ymax(n_j)]$ and $z_p\in[\zmin(n_k),\zmax(n_k)]$, respectively.
The neighboring points of point $q$ is simply the intersection of the three corresponding point sets: $\nb(q)=\ps_i^x \cap \ps_j^y \cap \ps_k^z$. Point correspondence can also be found efficiently within the neighborhood point set by using a distance threshold. And the adjacent intervals $\cup\nb(q)$ around $\nb(q)$ can be retrieved in a similar fashion.


\paragraph{Local per-point octrees}
Although the global coordinate interval tree can be used to find a local neighborhood for any given point, the neighborhood is an merely unstructured point set. To conduct point convolution, a distance metric between points is required to apply convolutional operations with distance-based kernels~\cite{qi2017pointnet++,li2018pointcnn}. To this end, we need to sort the set of neighboring points into a structured organization based on surface-aware metric. This is achieved by maintaining per-point octrees so that the surface-aware neighborhood in arbitrary scale can be found efficiently.


Given a point $p \in \ps$, we first retrieve its local neighborhood $\nb(p)$ using the coordinate interval trees.
We then divide the extended point set $\cup\nb(p)$ and its according to the eight quadrants of the Cartesian coordinate system originated at $p$. Within each quadrant, we add the point that is the closest to $p$ as the child of the corresponding direction, if the shortest distance is smaller than a threshold $d^\text{T}$. If some quadrant does not contain a point, however, the corresponding child node is left empty. The detailed description of per-point octree construction can be found in the supplemental material. After this process, we compute for each point a 1-ring neighborhood organized in a \emph{direction-aware}.
Based on the direction-aware octrees, one can easily expend the 1-ring neighborhood of a point into multiple rings through chaining octree-based neighbor searches; see Figure~\ref{fig:datastructure}.

The neighborhood search enabled by the per-point octrees has two important characteristics.
\emph{First}, for each point, its eight neighbor points (child nodes) are scattered in the eight quadrants of its local Cartesian frame. When finding n-ring neighbors based on the octree-based point connections, the consecutive searches can roughly follow the eight directions. Consequently, the octrees direct the search to find evenly distributed neighbors in all directions, which can be expensive to realize with naive region growing.
\emph{Second}, since the octrees maintain fine-scale local neighborhood, the search path along the octree-based connections approximately follows the 3D surface. This results in surface-aware n-ring neighborhoods.
Both the two characteristics benefit substantially point convolution in learning improved 3D features over those working with Euclidean-distance-based neighborhoods as demonstrated in \Sec{result}.


\subsection{Fusion-aware Point Convolution}
\label{sec:conv}

We propose a convolution operation which extends a recent work PointConv\cite{wu2019pointconv} with intra-frame and inter-frame feature fusion named fusion-aware point convolution. PointConv introduces a novel convolution operation over point cloud:
\begin{equation*}
\text{PC}_{p}(W,F)=\sum_{\Delta p \in \Omega}W(\Delta p)F(p+\Delta p),
\end{equation*}
where 
$F(p+\Delta p)$ is the feature of a point in the local region $\Omega$ centered at $p$ and $W$ is weight function.

RGB-D frame sequence is a mixed data of rich 2D and 3D information with time stamp. However, PointConv only utilize limited 3D information and the rest does not make contribution to the segmentation task. To improve this method, there are three primary questions that we seek to answer within the section. First, PointConv is mainly about 3D, but how to fuse 2D information properly with 3D? Second, can we construct better local area $\Omega$ which ensures the neighborhood would be more relevant? Thirdly, how to utilize the inter-frame information given by the sequence?


\paragraph{2D-3D feature fusion}
\label{2d3d}
3D feature fusion with online RGB-D reconstruction should better utilize the temporal correlation between adjacent frames, which goes beyond simplistic projection-based 2D-3D correlation.

The feature encoding at a 3D point should consider all the matched pixels in different frames if it is observed from multiple views. Pixel correspondence between consecutive frames can be easily retrieved based on $\gtx$, $\gty$ and $\gtz$.
This way, each 3D point $p$ in the scene would have a set of corresponding 2D pixels $I(p)=\{c^k | k\in n\}$. We can extract feature for each pixel $c^k$ intra-frame via 2D convolution on image. We adopt a pre-trained $\text{FuseNet}$\cite{Hazirbas2016FuseNet} without multi-scale layers as our 2D feature encoder (\Sec{network}):
\begin{equation*}
F^{2D}(c^k)=\text{FuseNet}(f_k,c^k),
\end{equation*}
where $\text{FuseNet}(f_k,c^k)$ is the 2D feature given by $\text{FuseNet}$ for the pixel $c^k$ in frame $f_k$. Therefore, each 3D point $p$ in the scene has a set of corresponding 2D features and max-pooling is adopted to fuse them into one feature:
\begin{equation*}
F^{\text{2D3D}}(p)=\text{maxpooling}\{F^{2D}(c^k)| c^k\in I(p)\}
\end{equation*}




\paragraph{Octree-induced surface-aware 3D convolution}
\label{geofusion}
Most deep convolutional neural networks for 3D point clouds gather neighborhood information on the basis of Euclidean distance. Apparently, geodesic distance can better capture the underlying geometry and topology of 3D surfaces. We propose octree-based neighborhood to take advantage of the geodesic approximation offered by our octree structure.

In particular, the local region $\Omega$ for each point $p$ is given by its octree $\lt(p)$. This approach could ensure the neighborhood would only enlarge along the object surface which is surface-aware, but not skip some gaps to reach shortest distance on Euclidean metric. A visual example is shown in Figure\ref{fig:geodesic}, the neighborhood searched by our approach would be more semantic related with the central point. This characteristic would benefit the following segmentation task. The construction of n-ring local region $\Omega^n(p)$ follows:
\begin{equation*}
\begin{aligned}
\Omega^n(p) &= \{\lt(p)|p\in O(\Omega^{n-1}(p))\},\\
\Omega^1(p) &= \Omega(p).
\end{aligned}
\end{equation*}
We name the n-ring local region found $G^n(p^v)$ as octree based neighborhood, and we can adopt it to improve our fusion-aware point convolution. The convolution formulation $\text{FPC}_{p}(W,F^{\text{2D3D}})$ follows as below:
\begin{equation*}
\begin{aligned}
\text{FPC}_{p}(W,F^{\text{2D3D}})= \sum_{\tiny \mathclap \Delta p \in \Omega^n(p)}W(\Delta p)F^{\text{2D3D}}(p+\Delta p).
\end{aligned}
\end{equation*}

\begin{figure*}[t]
\centering
\begin{overpic}
[width=0.8\linewidth]
{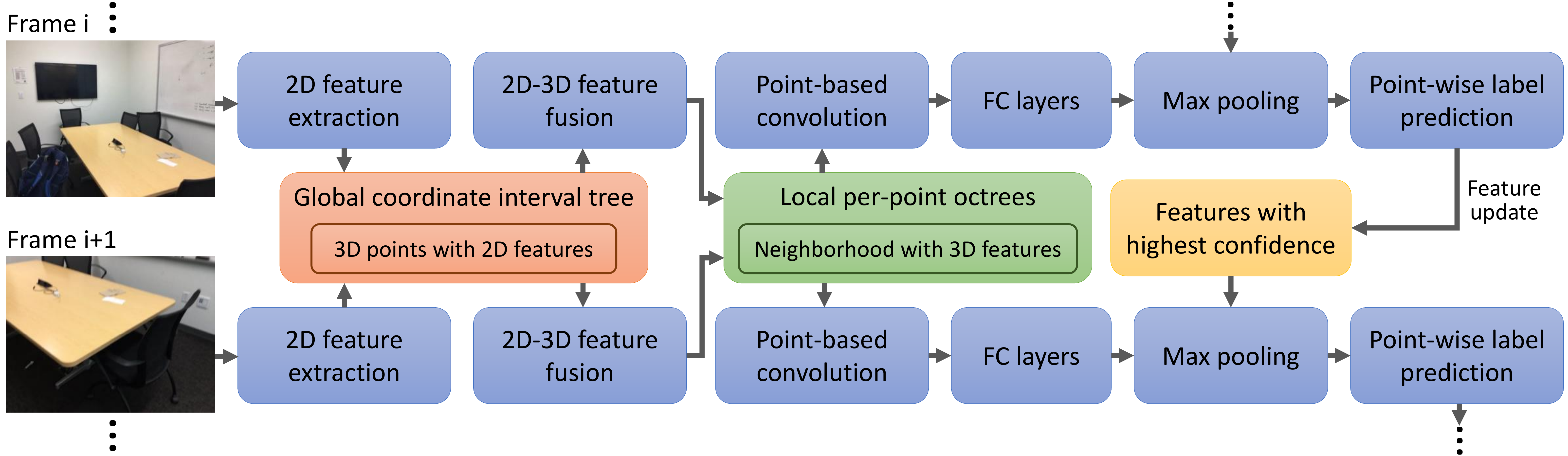}\myfigurename{}
\end{overpic}\vspace{-10pt}
\caption{
Network architecture. We show the pipeline with two consecutive frames. The global and local trees are dynamic data structure which evolves through time. The network output point-wise labels along with confidence. The labeling prediction is used to update features from the previous frame to be fused into the next frame.
}
\label{fig:network}\vspace{-12pt}
\end{figure*}

\paragraph{Frame-to-frame feature fusion}
\label{f2f}
Beside 2D-3D feature fusion, inter-frame information about segmentation uncertainty can benefit the segmentation task as well. A recent work about active scene segmentation \cite{zheng2019active} demonstrate that the segmentation entropy or uncertainty is crucial for online processing. We introduce a frame-to-frame feature fusion which utilize the segmentation results given by previous frames to improve the performance for following frames.

For each 3D point $p$, our method would update its segmentation result if it is observed by a new frame $f_i$. Although we do not know the result is correct or not in the test, predicted segmentation uncertainty $U(p,i)$ given by the convoluted feature $\text{FPC}^i_{p}(W,F^{\text{2D3D}})$ at frame $i$ can be easily retrieved by the network. Our basic idea here is that if $p$ has low segmentation uncertainty in frame $f_i$, the current form of feature fusion should be useful in the future prediction.

In practice, we record every uncertainty $U(p,i)$ when processing the frame sequence. Further, our method conducts maxpooling operation on $\text{FPC}^{\text{current}}_{p}(W,F^{\text{2D3D}})$ which was updated in the current frame with feature $\text{FPC}^i_{p}(W,F^{\text{2D3D}})$.
Thus, we rewrite the feature fusion as follows:
\begin{equation*}
\begin{aligned}
&F^{\text{fused}}(p) = \\ &\text{maxpool}\{\text{FPC}^{\text{current}}_{p}(W,F^{\text{2D3D}}),\argmin_{\text{FPC}^i_{p}}U(p,i)(W,F^{\text{2D3D}})\}
\end{aligned}
\end{equation*}
This way, the intra- and inter-frame information of $p$ are fused into $F^{\text{fused}}(p)$, maximally utilizing the information of the input RGB-D frames and significantly improving the performance of semantic segmentation.

\subsection{Online Segmentation Network}
\label{sec:network}

As shown in Figure~\ref{fig:network}, the the backbone of our proposed online segmentation network is the global-local tree structure we introduced in \Sec{tree}. The global coordinates interval tree helps mapping corresponding 2D features for 2D-3D feature fusion and the local octrees help searching surface-aware neighborhood for 3D fusion-aware convolution.


\paragraph{Network architecture}
The backbone of our 2D feature encoder is $\text{FuseNet}$\cite{Hazirbas2016FuseNet}. 
Note, however, we discard the multi-scale layers in our implementation since the re-sampling operation is too time-consuming. Please refer to the supplemental material for details. This modification enables our method to achieve a close-to-interactive performance ($10$ FPS) with high accuracy. 

According to \Sec{conv}, the 2D features $\text{FuseNet}(f_i,c^k)$ in different frames corresponding to the same 3D point $p$ are fused as $F^{\text{2D3D}}(p)$. Fused features are adopted in our proposed fusion-aware convolution, and the convoluted feature $\text{FPC}^i_{p}(W,F^{\text{2D3D}})$ for point $p$ in frame $i$ could be calculated based on its n-ring neighbors $\Omega^n(p)$ along geometry surface. The convoluted feature is sent to a simple fully connected network $FC$ which consists of 3 mlp layers to get the final feature which length is 128. It then be further fused with the selected feature with highest segmentation confidence in previous frames through max-pooling. Finally, we use it to predict the semantic label for $p$ with a one layer classifier. Note that, our network updates $\argmin_{\text{FPC}^i_{p}(W,F^{\text{2D3D}})}U(p,i)$ simultaneously to ensure the feature with the lowest segmentation uncertainty would be adopted in the future fusion-aware processing.

\paragraph{Training details}
The batch size in training is 64. For each batch, we randomly select 8 different scenes which each contributes a sequence of 8 frames. The first frame of each sampled sequences are the first 8 data in each batch and the $n$-th frames are the $(8*(n-1))$-th to $(8*n-1)$-th data in each batch. We back-propagate the training gradients once every 8 forward passes of frames, and update the network weights after the forward pass of the whole batch. The training of our network on ScanNet~\cite{dai2017scannet} takes about $40$\textapprx$48$ hours on a single Titan Xp GPU. For more details please refer to the supplemental material.


\section{Results and evaluations}
\label{sec:result}
We first introduce our benchmark dataset and how we setup our experiments. Comparisons with some state-of-the-art alternatives are presented on both online and offline semantic segmentation tasks. We then conduct extensive evaluation for each components of our method. We also demonstrate the advantage of surface-aware characteristic of our method through some experiments.

\subsection{Benchmarks}

\paragraph{Dataset} We evaluate our method on two datasets: ScanNet\cite{dai2017scannet} and SceneNN\cite{scenenn-3dv16}. ScanNet contains 1513 scanned scene sequences, out of which we use 1200 sequences for training and the rest 312 for testing. SceneNN contains 50 high-quality scanned scene sequences with semantic label. However, this dataset is not specifically organized for online segmentation task. Some of the scanned scene sequences do not have camera pose information. Color image and depth map are not well aligned in some of the sequences as well. After some filtering work, we select 15 clean sequences from SceneNN with proper scanned information for our evaluation.

\paragraph{Experiment configuration} To evaluate the performance of our method, we adopt accuracy and IOU as two indicators in our experiments. Since different online segmentation methods may adopt different 3D reconstruction approaches, it is really difficult to measure these two indicators in different 3D point clouds. In our experiment, we project the semantic labels of 3D points into their corresponding 2D frames and measure the accuracy and IOU in 2D.

\begin{table}[t!]
	\centering
	\caption{Accuracy comparison between our method and two state-of-the-art online scene segmentation methods.}\vspace{-6pt}
	\scalebox{0.9}{\setlength{\tabcolsep}{1.2mm}{
	\begin{tabular}{@{}lllll@{}}
		\toprule
		Dataset & SemanticFusion\cite{mccormac2017semanticfusion} & ProgressiveFusion\cite{pham2019real} & Ours \\ \midrule
		ScanNet & 0.518 & 0.566	& \textbf{0.764} \\
		SceneNN & 0.628	& 0.666 & \textbf{0.675}
		\\ \bottomrule
	\end{tabular}}}\vspace{-10pt}
	\label{tab:online}
\end{table} 

\subsection{Semantic Segmentation Comparison}
\label{subsec:result-seg}

\paragraph{Comparison with other online methods} Our method is compared to two state-of-the-art online segmentation methods for indoor scenes:  SemanticFusion\cite{mccormac2017semanticfusion} and ProgressiveFusion\cite{pham2019real}. The comparison is conducted on SceneNN and ScanNet respectively. The mean accuracy of three methods are shown in Table~\ref{tab:online}. From the results we can clearly see that our method gets the highest segmentation accuracy on both datasets. Note that we only train our method on ScanNet dataset and do not fine-tune it on SceneNN, and our method still outperform other methods on both datasets. This result demonstrate the generality of our method and it can be easily adopted on different dataset.

\vspace{-2pt}
\begin{table*}[t]
	\centering
	\caption{IOU comparison between our method and state-of-the-art offline scene segmentation methods. Our method has the highest mean IOU, outperforming the state-of-the-art methods for nine semantic categories.}\vspace{-6pt}
	\scalebox{0.72}{\setlength{\tabcolsep}{0.6mm}{
	\begin{tabular}{@{}l|c|c|c|c|c|c|c|c|c|c|c|c|c|c|c|c|c|c|c|c|c@{}}
		\toprule
		model         & \textbf{mean}   & wall   & floor & cabinet & bed   & chair & sofa  & table & door  & window & bookshelf & picture & counter & desk  & curtain & fridge & bathshade & toilet & sink  & bathtub & others \\ \midrule
		SparseConvNet & 0.685 & 0.828  & \textbf{0.950}  & 0.620   & 0.805 & 0.894 & 0.825 & 0.707 & 0.633 & 0.588  & 0.788     & 0.252   & 0.601   & 0.592 & 0.681   & 0.428        & 0.607          & \textbf{0.928}  & 0.596 & 0.881   & 0.504          \\
		MinkowskiNet  & 0.715  & 0.841 & 0.949 & \textbf{0.641}  & 0.806 & \textbf{0.900}   & \textbf{0.845} & \textbf{0.745}& 0.648 & 0.608  & \textbf{0.792}     & 0.289   & \textbf{0.637}   & \textbf{0.65}  & 0.742   & 0.509        & 0.690           & 0.916  & \textbf{0.689} & 0.832   & \textbf{0.570}           \\
		PointConv     & 0.580  & 0.741   & 0.948 & 0.474  & 0.672 & 0.813 & 0.633 & 0.651 & 0.346 & 0.446  & 0.713     & 0.067   & 0.568   & 0.525 & 0.551   & 0.370        & 0.520          & 0.840  & 0.590 & 0.750   & 0.387          \\
		Ours          & \textbf{0.720}  & \textbf{0.862}  & 0.924 & 0.615  & \textbf{0.848} & 0.716 & 0.804 & 0.637 & \textbf{0.680} & \textbf{0.698}  & 0.724     & \textbf{0.513}   & 0.617   & 0.588 & \textbf{0.764}   & \textbf{0.734}         & \textbf{0.696}           & 0.870   & 0.681 & \textbf{0.885}   & 0.556          \\ \bottomrule
	\end{tabular}}}
	\label{tab:offline}\vspace{-8pt}
\end{table*} 
\vspace{-2pt}

\begin{figure*}[!t]
\centering
\begin{overpic}
[width=\linewidth]
{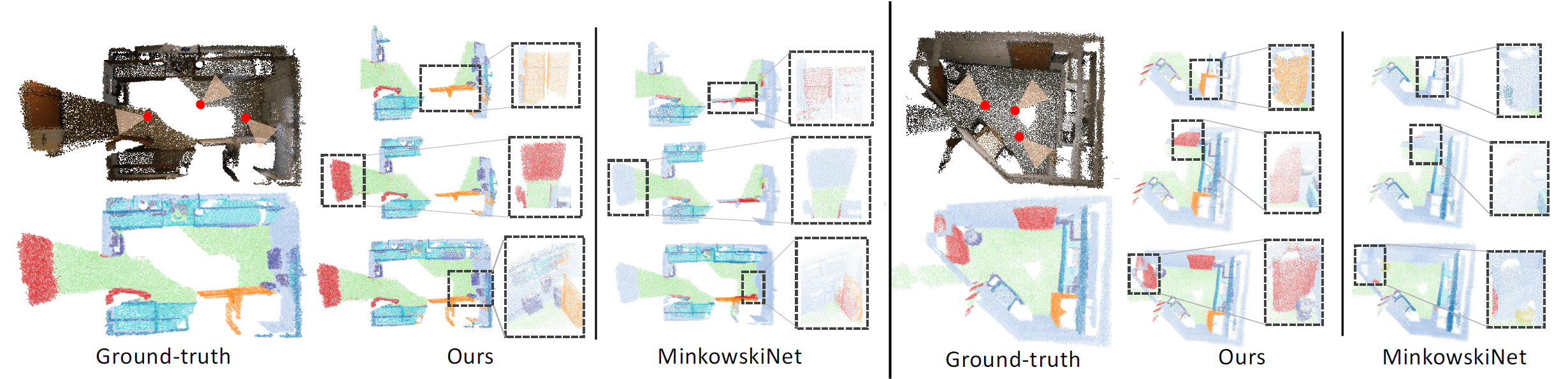}\myfigurename{}
\end{overpic}
\caption{
Visual comparison between our method and MinkowskiNet~\cite{choy20194d}. Our method works better than MinkowskiNet especially on those small and incomplete objects. Live demo is provided in the accompanying video.
}
\label{fig:comparison}\vspace{-12pt}
\end{figure*}

\paragraph{Comparison with offline methods} To further demonstrate the superiority of our segmentation method, we setup a comparison with three state-of-the-art offline segmentation methods (SparseConvNet\cite{3DSemanticSConvNet}, PointConv\cite{wu2019pointconv} and MinkowskiNet\cite{choy20194d}) for indoor scenes as well. Note that, there is a challenge in online segmentation methods when compare to offline alternatives. Partial scene would be more difficult to be segmented than the whole scene. We show some visual comparison on partial scenes in Figure~\ref{fig:comparison}. Our method can achieve much better results on these challenging cases which is very crucial for online tasks. We also present a segmentation comparison on the complete scene in Table~\ref{tab:offline}. Our method achieves a comparable performance with the offline alternatives.

\subsection{Ablation study}

\begin{figure}[t]
\centering
\begin{overpic}
[width=\linewidth]
{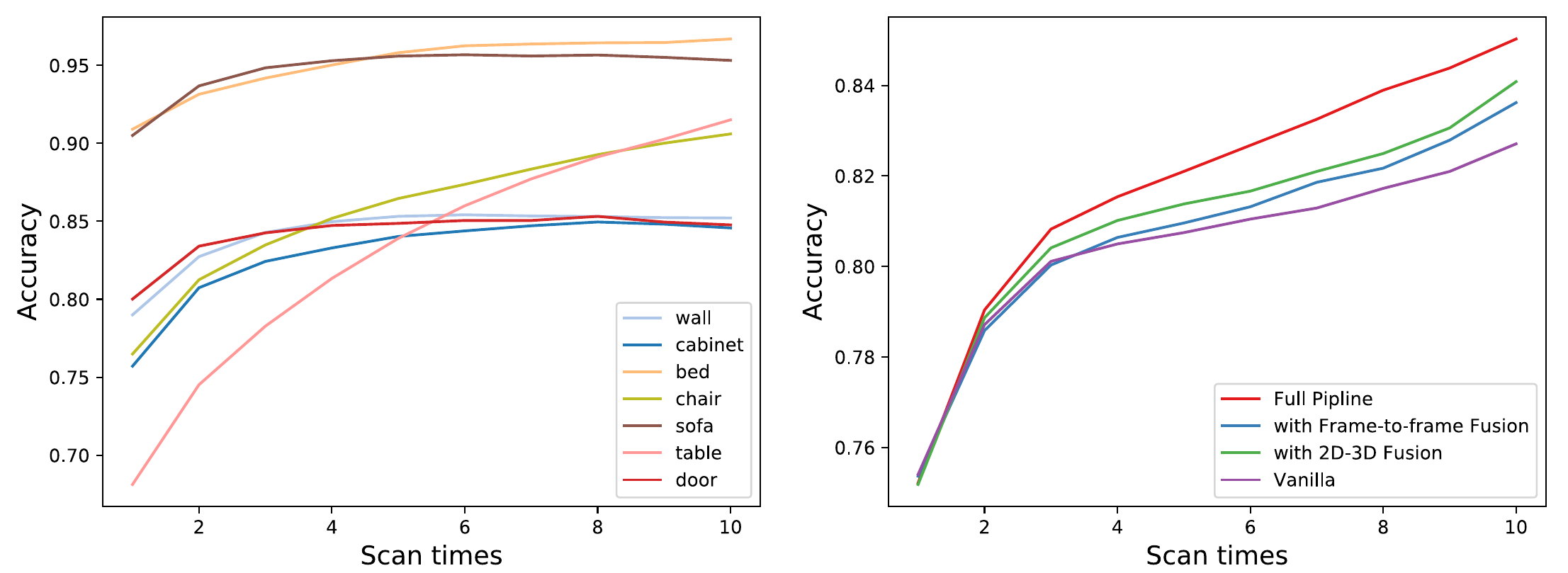}\myfigurename{}
\end{overpic}
\caption{
Segmentation accuracy improves over time. The left plot shows per-category accuracy. Objects with complex structures, such as tables and chairs, benefit more from our feature fusion. The right is mean accuracy. Segmentation accuracy improves significantly with increasing scans.
}
\label{fig:scantimes}\vspace{-10pt}
\end{figure}

\paragraph{Feature fusion study}
We investigate the effect of some crucial designs on segmentation performance. We turn off the 2D-3D feature fusion and frame-to-frame feature fusion in succession to assess how these two components would benefit our method. The results are shown in Table~\ref{tab:ablation}. Without 2D-3D feature fusion, our method cannot improve performance anymore with multi-view information in the sequence. The absence of frame-to-frame feature fusion makes our method lose the ability of learning from history in the sequence. We observe significant drop on performance if we turn off these designs, which prove the importance of these two designs. Similar results are also demonstrated in the right plot of Figure~\ref{fig:scantimes}.


\begin{table}[t!]
	\caption{Ablation study on the two feature fusion operations of our method. The results justify each design.}\label{tab:ablation}\vspace{-6pt}
	\begin{center}
		\scalebox{0.9}{
			\begin{tabular}{c|c|c}
				\whline{1.15pt}
				{2D3D feature} & {frame-to-frame feature} & {mean IOU}\\
				\whline{0.65pt}
				$\times$&  $\times$&  $0.711$\\
				\checkmark & $\times$ & $0.718$\\
				$\times$& \checkmark & $0.713$\\
				\checkmark & \checkmark & $\bf{0.720}$ \\
				\whline{1.15pt}
		\end{tabular}}\vspace{-6pt}
	\end{center}
\end{table} 

\paragraph{Temporal information study} The segmentation label of a 3D point would be updated if new information is fused in the sequence. To further investigate how our fusion-aware point convolution benefit the segmentation performance, we plot the accuracy of point labeling with increasing feature fusion times in Figure~\ref{fig:scantimes}. In the plots, ``vanilla'' refers to our basic model with global-local tree and improved point convolution. We observe a clear increase of segmentation accuracy with increasing scans. Such effect is more observable on objects with complex structures, such as tables and chairs. Note that the performance of the vanilla model also improves with increasing scans. This is because both tree construction and point convolution benefit from a more complete point cloud.

\begin{figure}[t]
\centering
\begin{overpic}
[width=0.96\linewidth]
{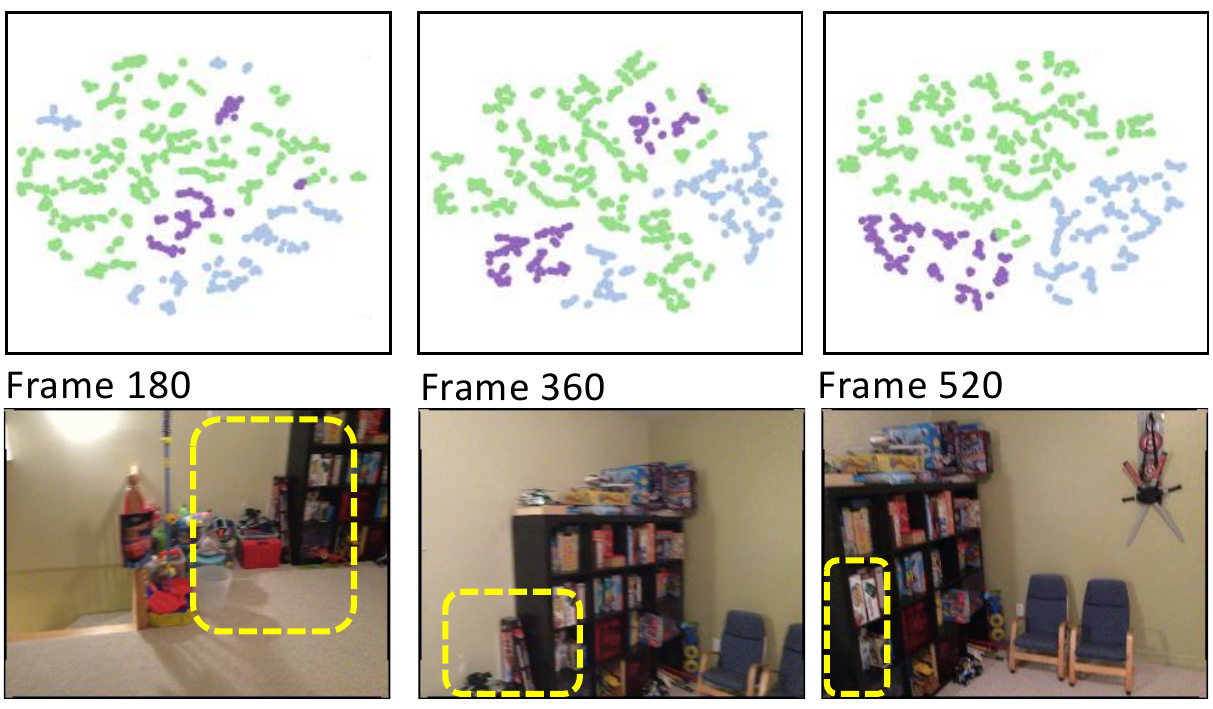}\myfigurename{}
\end{overpic}
\caption{
The evolution of point-wise feature embedding with more frames are acquired and fused. We show the t-SNE plots of the feature embedding of the same group of points with semantic labels indicated by color.
}
\label{fig:feaEmd}\vspace{-12pt}
\end{figure}

To explore why point with more fusion times would have better segmentation performance, we track feature changes of some points in Figure~\ref{fig:feaEmd}. We plot the feature embedding of the points marked in yellow box, and we found the boundary between points with different semantic labels are getting clear with during the scanning. Our fusion-aware features are more separable in the embedding space which make the segmentation network easier to train.

\begin{figure}[t]
\centering
\begin{overpic}
[width=\linewidth]
{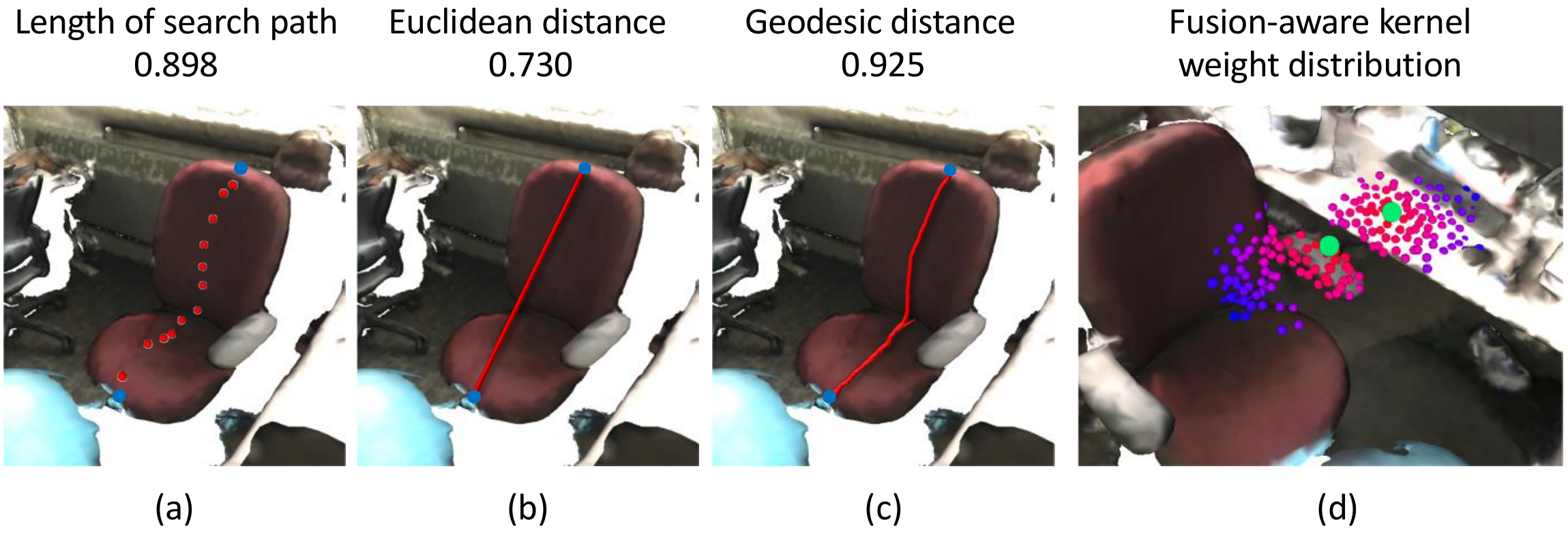}\myfigurename{}
\end{overpic}
\caption{
(a-c): Tracing path between two selected points along the local octrees. We compare the path length to the Euclidean distance and ground-truth geodesic distance. We find that the path traced along the local octrees is a good approximate of the geodesic distance. (d): The kernel weight distribution of two points (green dot) are visualized (red is high and blue is low).
}
\label{fig:geodesic}\vspace{-12pt}
\end{figure}

\paragraph{Surface-aware convolution context} Another important characteristic of our method is the neighborhood used in point convolution is surface-aware. This is enabled by our local octree structure. To verify this claim, we select two points on the point cloud and estimate their distance in Figure~\ref{fig:geodesic}. We observe that tracing along the local octrees leads to a path whose length is closer to the ground-truth geodesic distance than the Euclidean distance. The visual example also demonstrates that the tree-induced path traces along the surface, making the point neighborhood surface-aware. Figure~\ref{fig:geodesic} also visualizes the kernel weight
distribution of two points on the point cloud, showing that neighboring points which are geodesically closer have higher weights.

\begin{figure}[t]
\centering
\begin{overpic}
[width=\linewidth]
{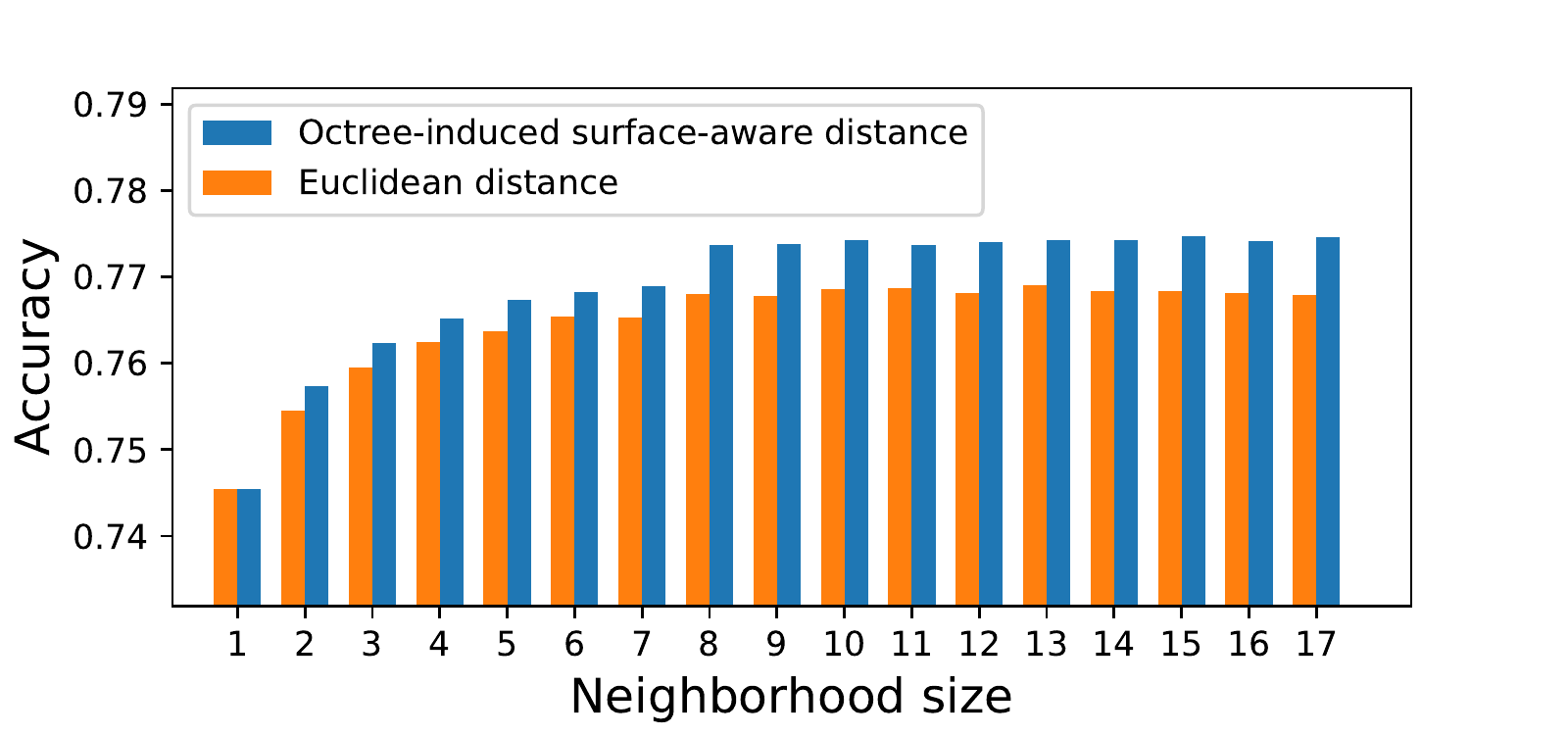}\myfigurename{}
\end{overpic}
\caption{
Segmentation accuracy of different neighborhood searching approaches. Our local-octree-induced neighborhood always leads to better segmentation accuracy than Euclidean based. We also find that increasing the neighborhood size more than a certain value would not improve the performance significantly.
}
\label{fig:octvseu}\vspace{-12pt}
\end{figure}

This characteristic makes our method more geometry-aware. To verify the geometric information would benefit the online segmentation, we design a comparison between convolutions based on Euclidean distance and our local-octree-induced distance in Figure~\ref{fig:octvseu}. The results show that our surface-aware convolution demonstrates a better performance all the way along the increase of convolution area (size of receptive field).

\section{Conclusion}
For the task of semantic segmentation of a 3D scene being reconstructed with RGB-D fusion, we have presented a tree-based dynamic data structure to organize online fused 3D point clouds. It supports 3D point convolution over time-varying geometry, fusing information between 2D and 3D and from frame to frame.
Our method achieves online segmentation at close-to-interactive frame-rate while reaching state-of-the-art accuracy.
Our current method has a few limitations on which we plan to investigate in future works.
\emph{First}, our current system does not support data streaming, thus confining the per-frame point cloud density we could handle due to memory limit.
\emph{Second}, our method still relies on accurate camera poses for high-quality point fusion and convolution. Although it works well with the popular RGB-D fusion methods (such as BundleFusion~\cite{Dai2017}), it is interesting to investigate how to integrate online semantic segmentation with camera tracking, accomplishing semantic SLAM with a state-of-the-art accuracy for both. Further, we would like to test our method on outdoor scenes~\cite{SYNTHIA}.

\section*{Acknowledgement}
We thank the anonymous reviewers for the valuable suggestions.
We are also grateful to Quang-Hieu Pham for the help on comparison with ProgressiveFusion.
This work was supported in part by NSFC (61572507, 61532003, 61622212), NUDT Research Grants(No.ZK19-30) and Natural Science Foundation of Hunan
Province for Distinguished Young Scientists (2017JJ1002).


{\small
\bibliographystyle{ieee_fullname}
\bibliography{fusionconv}
}
\clearpage
\appendix

\section{Supplementary Material Introduction}
This supplemental material contains four parts:
\begin{itemize}
	\item Section~\ref{sec:AlgDetails} reports the construction details of global coordinate interval tree and local octrees.
	\item Section~\ref{sec:NetDetails} reports the module details of our full pipeline network, which include parameter selection, network layers and loss functions.
	\item {\color{black}Section~\ref{sec:benchmarkresult} shows the comparison of the online methods\cite{narita2019panopticfusion}.}
	\item Section~\ref{sec:visresult} shows more progressive results of online semantic segmentation on ScanNet\cite{Dai2017} dataset.
\end{itemize}

\section{Algorithm Details}
\label{sec:AlgDetails}

Algorithm~\ref{algo:add} demonstrates the details about how we maintain nodes in a global coordinate interval tree when a point $p$ is detected in the coming new frame.

Algorithm~\ref{algo:oct} describes the construction of octree $\lt(p)$ for a given 3D point $p$ based on global coordinate interval trees  $\gtx$, $\gty$ and $\gtz$.

\begin{figure*}[t]
	\centering
	\begin{overpic}
		[width=\linewidth]
		{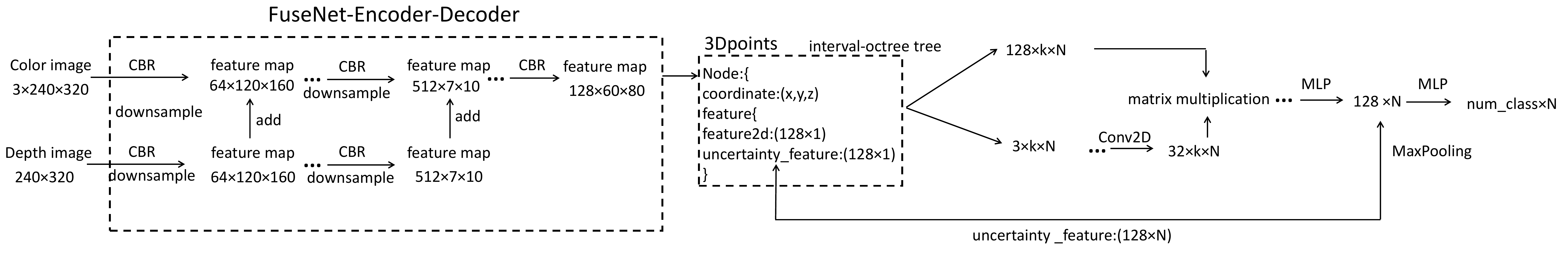}\myfigurename{}
	\end{overpic}
	\vspace{-12pt}
	\caption{
		Network architecture. 
	}
	\label{achi}\vspace{-12pt}
\end{figure*}

\begin{table*}[tp]
	\centering
	\caption{{\color{black}IOU comparison between our method and state-of-the-art online scene segmentation methods. Our method outperforms the state-of-the-art methods in 19 semantic categories(except curtain).}}\vspace{-6pt}
	\scalebox{0.75}{\setlength{\tabcolsep}{0.4mm}{
	\begin{tabular}{@{}llllllllllllllllllllll@{}}
		\toprule
		model          &\textbf{mean}   & wall   & floor & cabinet & bed   & chair & sofa  & table & door  & window & bookshelf & picture & counter & desk  &  curtain & fridge & curtain & toilet & sink  & bathtub & others \\ \midrule
        PanopticFusion & 0.529  & 0.491  & 0.688 & 0.604  & 0.386 & 0.632 & 0.225 & 0.705 & 0.434 & 0.293  & 0.815     & 0.348   & 0.241   & 0.499 & 0.669   & 0.507         & 0.649           & 0.442   & 0.796 & 0.602   & 0.561         \\ 
        Ours           & \textbf{0.630}  & 0.604  & 0.741 & 0.766  & 0.590 & 0.747 & 0.501 & 0.734 & 0.503 & 0.527  & 0.919     & 0.454   & 0.323   & 0.550 & 0.420   & 0.678         & 0.688         & 0.544  & 0.896 & 0.795   & 0.627    \\     \bottomrule
    \end{tabular}}}
	\label{tab:online}\vspace{-8pt}
\end{table*} 

\IncMargin{0.5em}
\begin{algorithm}
	\scriptsize
	\caption{Adding $p$ into global coordinate interval tree $\gtx$}
	\label{algo:add}
	\SetCommentSty{textsf}
	\SetKwInOut{AlgoInput}{Input}
	\SetKwInOut{AlgoOutput}{Output}
	\SetKwFunction{CreateNewNode}{CreateNewNode}
	\SetKwFunction{GetInterval}{GetInterval}
	\SetKwFunction{TraverseTree}{TraverseTree}
	\SetKwFunction{NearestNode}{NearestNode}
	\SetKwFunction{AddIntoNode}{AddIntoNode}
	\SetKwFunction{SetAsChild}{SetAsChild}
	\Indm
	\Indp
	\AlgoInput{ Existed nodes $n^i\in \gtx$, new 3D position $p=(x_{p},y_{p},z_{p})$ and threshold $d$.}
	\AlgoOutput{ Updated $\gtx$ and the a node $n$ contains $p$.}
	\tcp{find node n satisfy $\xmin(n)<x_{p}<\xmax(n)$}
	$n\leftarrow$\TraverseTree{$\gtx,x_{p}$}\;
	\If{$n$ satisfy $\xmin(n)<x_{p}<\xmax(n)$ exists}{$n\leftarrow$\AddIntoNode{$n,x_{p}$}\;\Return $\gtx,n$}
	\tcp{create a node $n$ contains $p$}
	\tcp{can pre-create neighbor nodes to reduce creation costs}
	$n\leftarrow$\CreateNewNode{$x_{p}$}\;
	$interval\leftarrow$\GetInterval{$x_{p},d$}\;
	$\xmin(n)\leftarrow interval_{min}$\;
	$\xmax(n)\leftarrow interval_{max}$\;
	\tcp{Find a nearest node in $R_x$ for n}
	$dist, node\leftarrow$\NearestNode{$n,\gtx$}\;
	\SetAsChild{$node,n$}\;\tcp{Rebalance with red-black tree}
	\Return $n,\gtx$\;
\end{algorithm}
\DecMargin{0.5em}

\IncMargin{0.5em}
\begin{algorithm}
	\scriptsize
	\caption{Construct octree $\lt(p)$ for $p \in \ps$}
	\label{algo:oct}
	\SetCommentSty{textsf}
	\SetKwInOut{AlgoInput}{Input}
	\SetKwInOut{AlgoOutput}{Output}
	\SetKwFunction{AdjIntervals}{AdjIntervals}
	\SetKwFunction{NearestDist}{NearestDist}
	\SetKwFunction{Dist}{Dist}
	\SetKwFunction{Initialize}{Initialize}
	\SetKwFunction{SameDirection}{SameDirection}
	\SetKwFunction{SetAsChild}{SetAsChild}
	\Indm
	\Indp
	\AlgoInput{ 3D point $p$, its neighborhood $\nb(p)$ and threshold $h=0.04$.}
	\AlgoOutput{ Octree $\lt(p)$.}
	\tcp{Initialize $\lt(p)$ with 8 nodes $\nb^m(p)$ at each direction $\overrightarrow{m}$}
	\Initialize( $\lt(p)$)\;
	\ForEach{$\nb^{\overrightarrow{m}}(p)$ in $\lt(p)$}
	{
		\NearestDist$(\nb^{\overrightarrow{m}}(p))=\inf$
	}
	\tcp{Extend the neighborhood for octree search}
	$\cup\nb(p)\leftarrow$\AdjIntervals($\nb(p)$)\;
	\ForEach{$p_i \in \cup\nb(p)$}
	{
		\tcp{Early finish}
		\If{\Dist$(p,p_i)<h$}
		{
			\Return$\lt(p_i)$\;
		}
		\ForEach{$\nb^{\overrightarrow{m}}(p)$ in $\lt(p)$}
		{
			\If{\SameDirection($\overrightarrow{m},\overrightarrow{p_ip}$)}
			{
				\If{\Dist$(p,p_i)<$\NearestDist$(\nb^{\overrightarrow{m}}(p))$}
				{
					NearestDist$(\nb^{\overrightarrow{m}}(p))$=\Dist$(p,p_i)$\;
					$\nb^{\overrightarrow{m}}(p)=p_i$\;
				}
			}
		}
	}
	\Return$\lt(p)$\;
\end{algorithm}
\DecMargin{0.5em}

\section{Network Architecture}
\label{sec:NetDetails}

Figure~\ref{achi} presents the detailed configuration of our whole network. There are two important difference between our method and other semantic segmentation network. First, since the predicted result in previous frames would be adopted in the future segmentation prediction, the online construction of our global-local tree would be time-consuming if no special design in training. We find that some parameter tuning would help us improve the efficiency. In training, we only take 4096 points for global-local tree construction in each frame and we find this configuration can take care both performance and efficiency. Secondly, our network need to take continuous frames as input since the frame-to-frame information is required during the training. However, adjacent frames may only have limited pixels with large differences. If we update the network weights after each frame forwarding in a continuous sequence, the training performance would be pretty bad since gradient generated in the loss backward may be too small to be stable for weight update. Therefore, we only update the network weights after 8 continuous loss backwards, and we find there is significant performance improvement when comparing to the naive weight update approach.    

\section{Website Benchmark Result}
\label{sec:benchmarkresult}
{\color{black}We test our result on the benchmark of the ScanNet Website. 
Because the reconstruction point cloud is different from the official point cloud, we have to map the labels to the nearest points. 
On Table~\ref{tab:online},  there is a decrease in the result due to the mismatch, but also outperform the state-of-the-art online method which proves the effectiveness of our methods.}


\section{More Online Segmentation Results}
\label{sec:visresult}
\begin{figure*}[]
	\centering
	\begin{overpic}
		[width=0.6\linewidth]
		{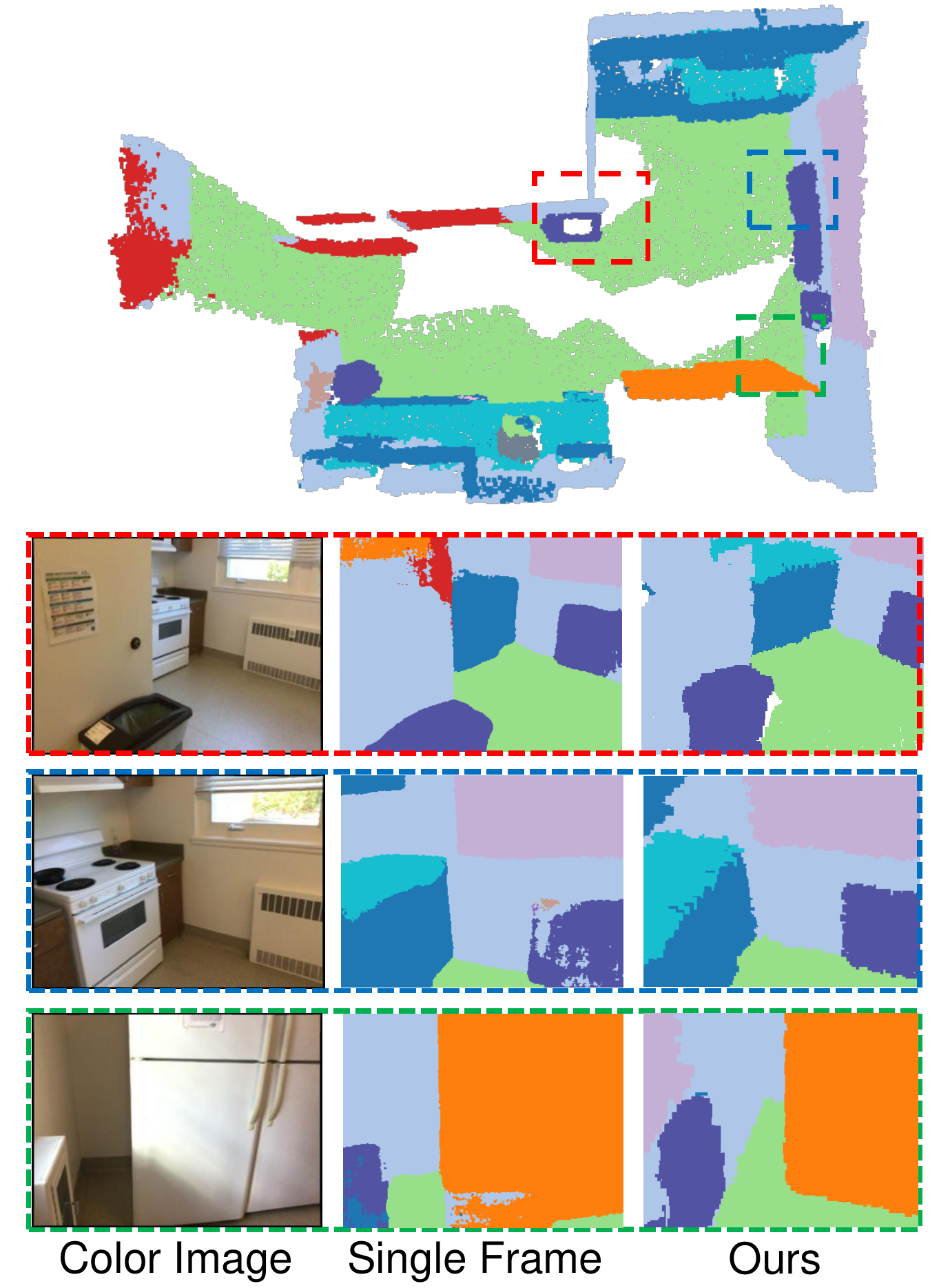}\myfigurename{}
	\end{overpic}\vspace{-6pt}
	\caption{
		{\color{black}The first row shows the complete result of our method. 
			The last three rows (in red, blue, gree boxes) give the input color image,
			semantic label of single frame and the projection result of our method.}
	}
	\label{fig:comparsion2d}\vspace{-16pt}
\end{figure*}

\begin{figure*}[]
	\centering
	\begin{overpic}
		[width=0.85\linewidth]
		{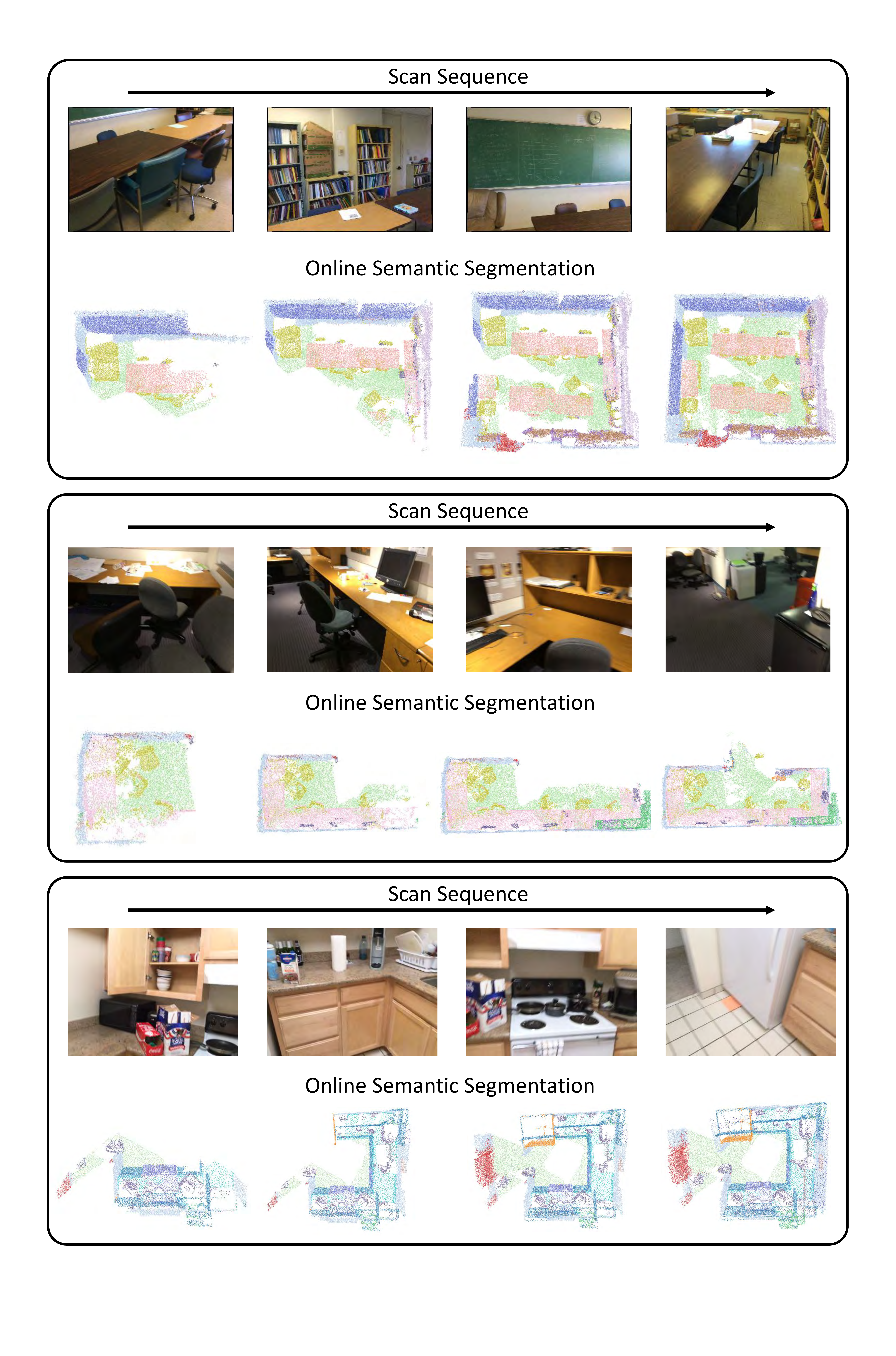}\myfigurename{}
	\end{overpic}
	\vspace{-12pt}
	\caption{
		Visual results of our online semantic segmentation method. Our method can work properly even the input scan is incomplete. 
	}
	\label{fig1}\vspace{-12pt}
\end{figure*}

\begin{figure*}[]
	\centering
	\begin{overpic}
		[width=0.85\linewidth]
		{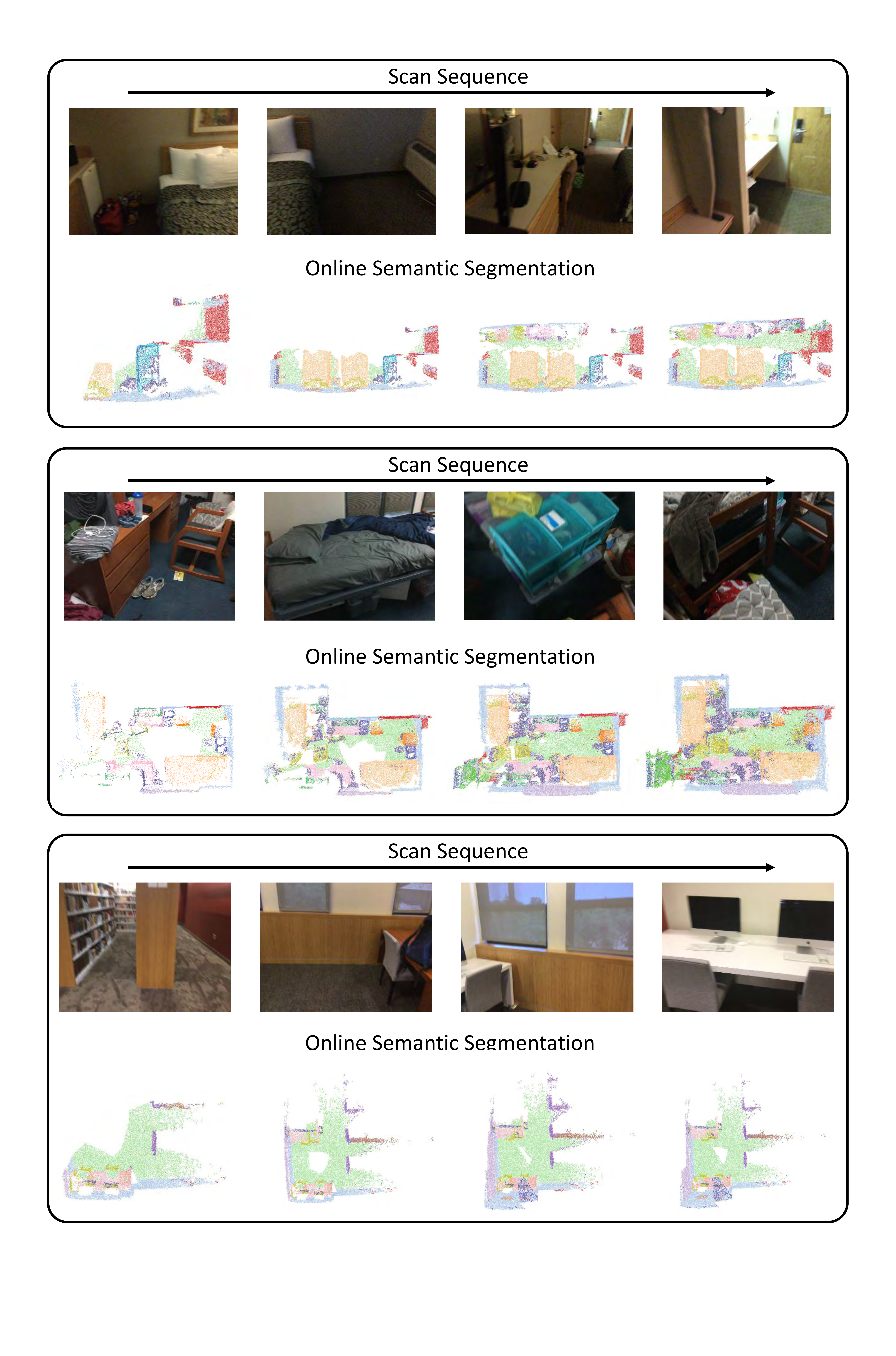}\myfigurename{}
	\end{overpic}
	\caption{
		Visual results of our online semantic segmentation method. Our method can work properly even the input scan is incomplete. 
	}
	\label{fig2}\vspace{-12pt}
\end{figure*}

\begin{figure*}[]
	\centering
	\begin{overpic}
		[width=0.85\linewidth]
		{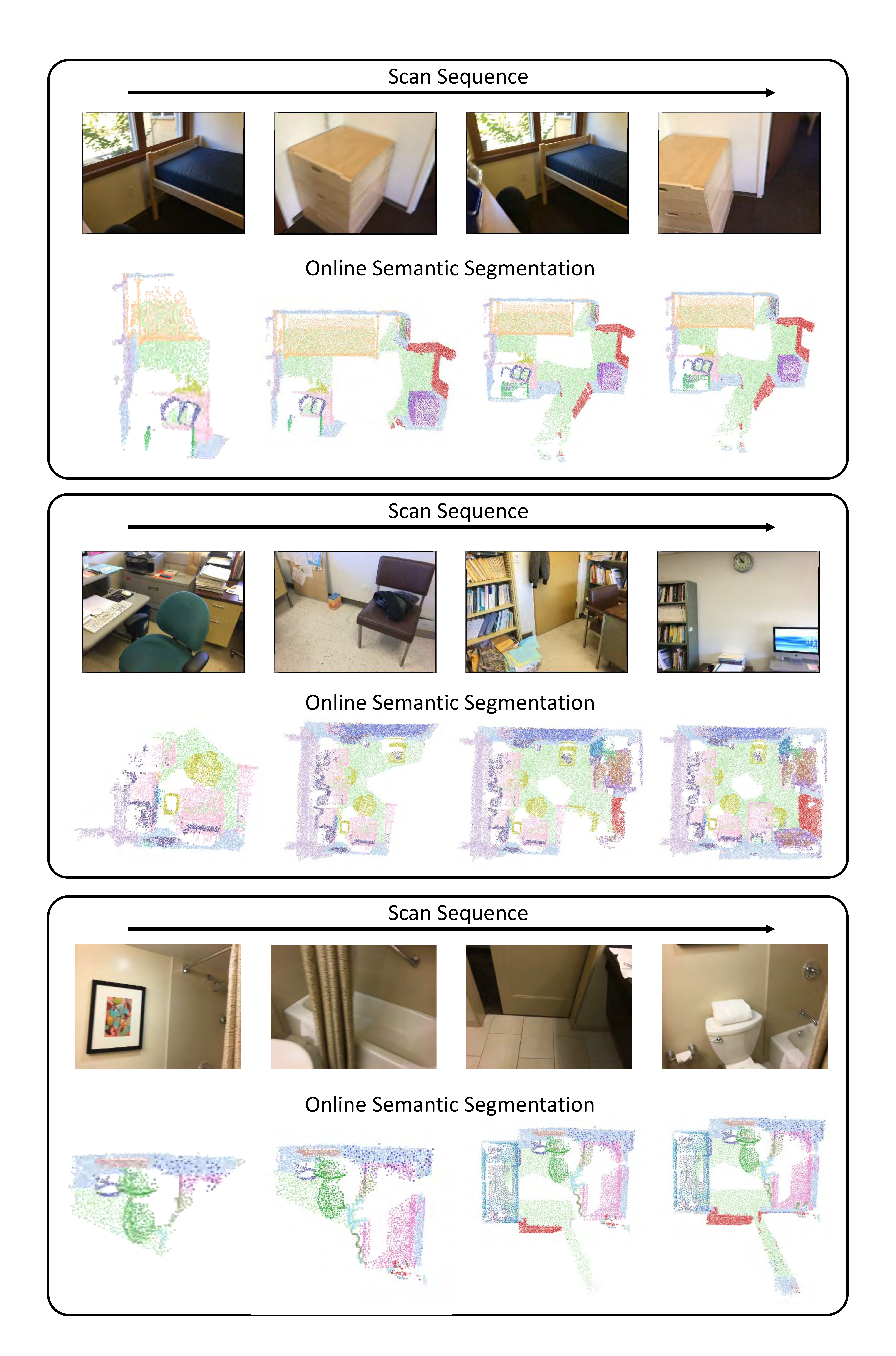}\myfigurename{}
	\end{overpic}
	\caption{
		Visual results of our online semantic segmentation method. Our method can work properly even the input scan is incomplete. 
	}
	\label{fig3}\vspace{-12pt}
\end{figure*}

\begin{figure*}[]
	\centering
	\begin{overpic}
		[width=0.85\linewidth]
		{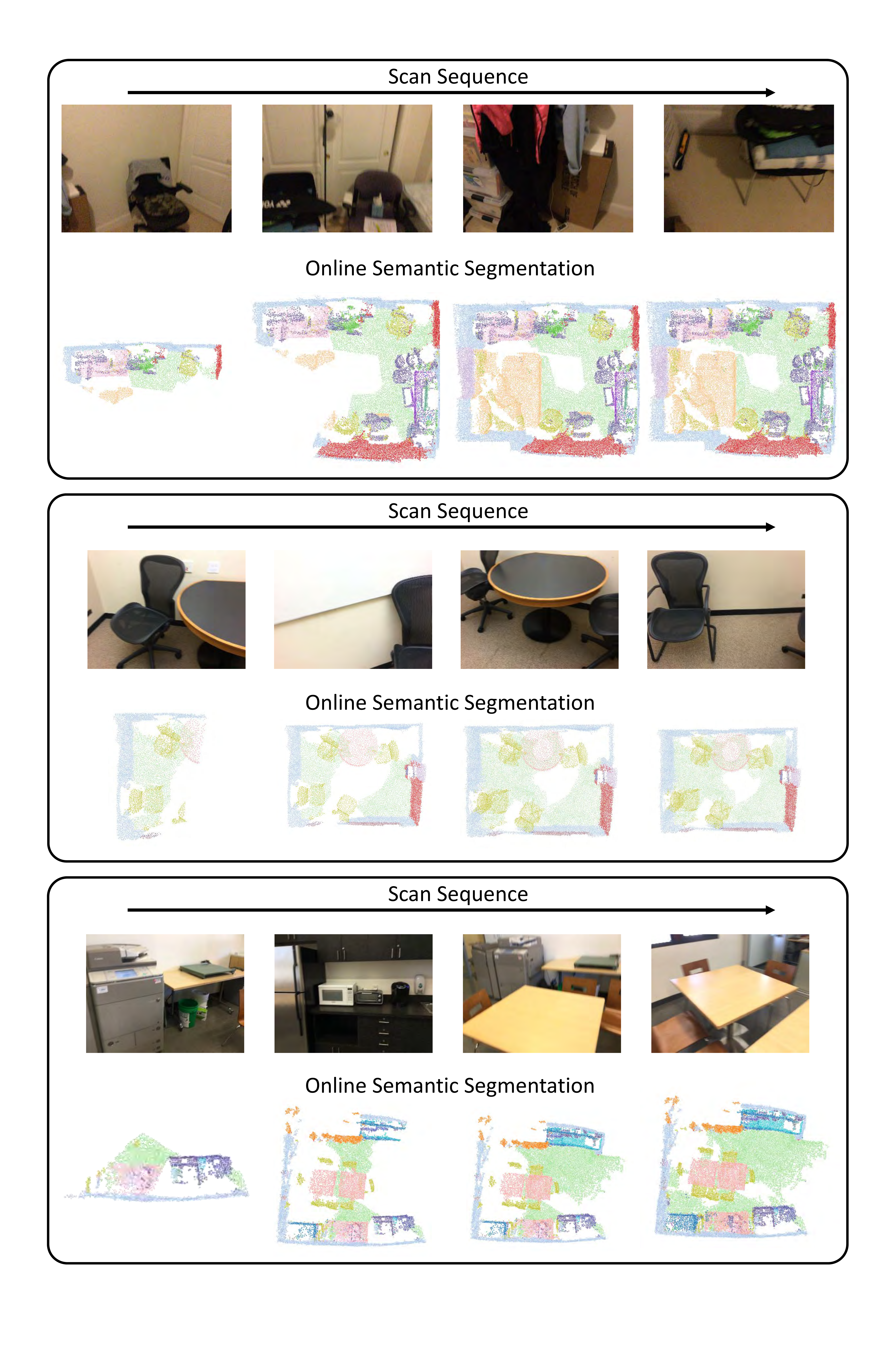}\myfigurename{}
	\end{overpic}
	\caption{
		Visual results of our online semantic segmentation method. Our method can work properly even the input scan is incomplete. 
	}
	\label{fig4}\vspace{-12pt}
\end{figure*}

\begin{figure*}[]
	\centering
	\begin{overpic}
		[width=0.85\linewidth]
		{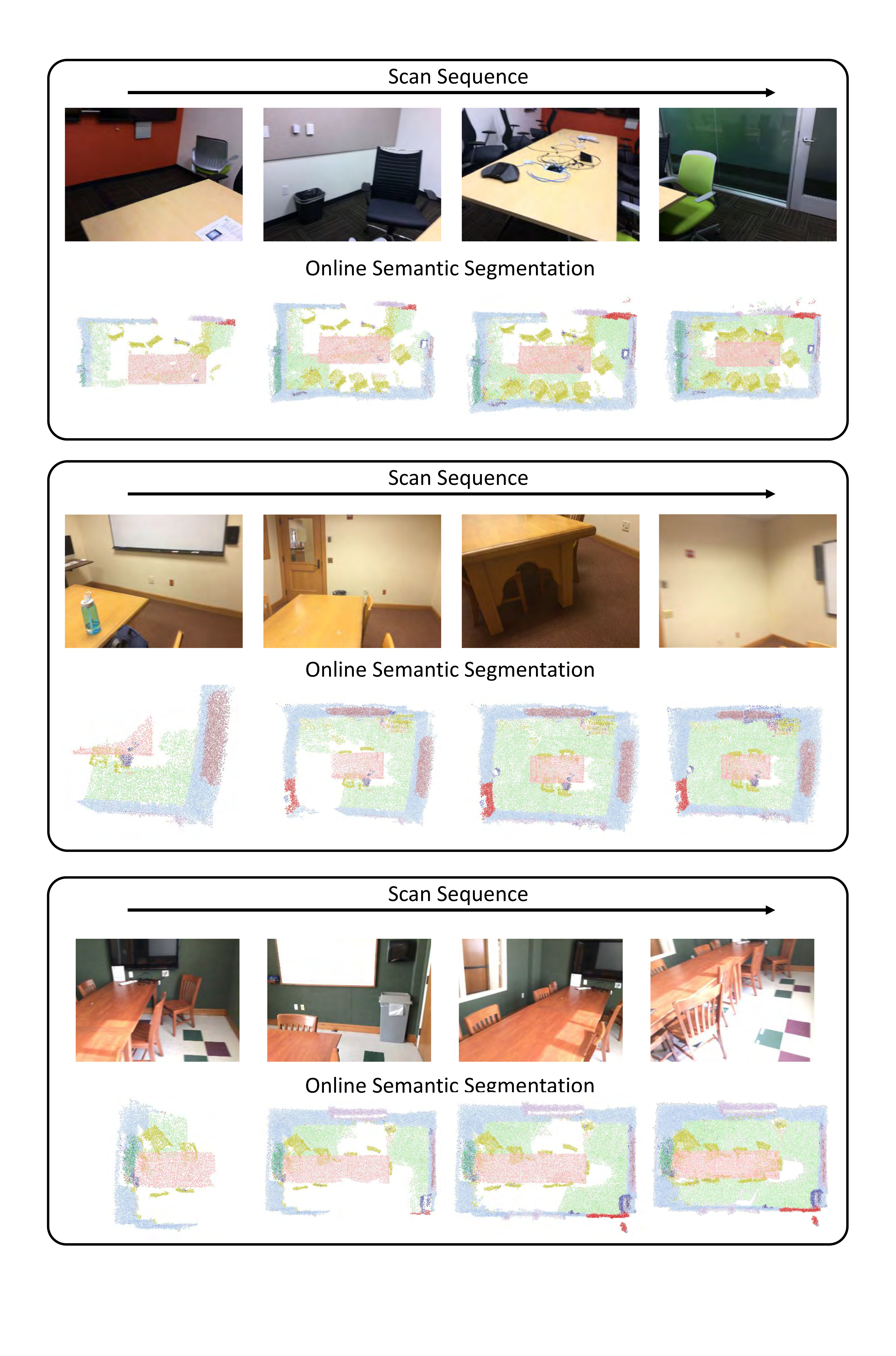}\myfigurename{}
	\end{overpic}
	\caption{
		Visual results of our online semantic segmentation method. Our method can work properly even the input scan is incomplete. 
	}
	\label{fig5}\vspace{-12pt}
\end{figure*}

\begin{figure*}[]
	\centering
	\begin{overpic}
		[width=0.85\linewidth]
		{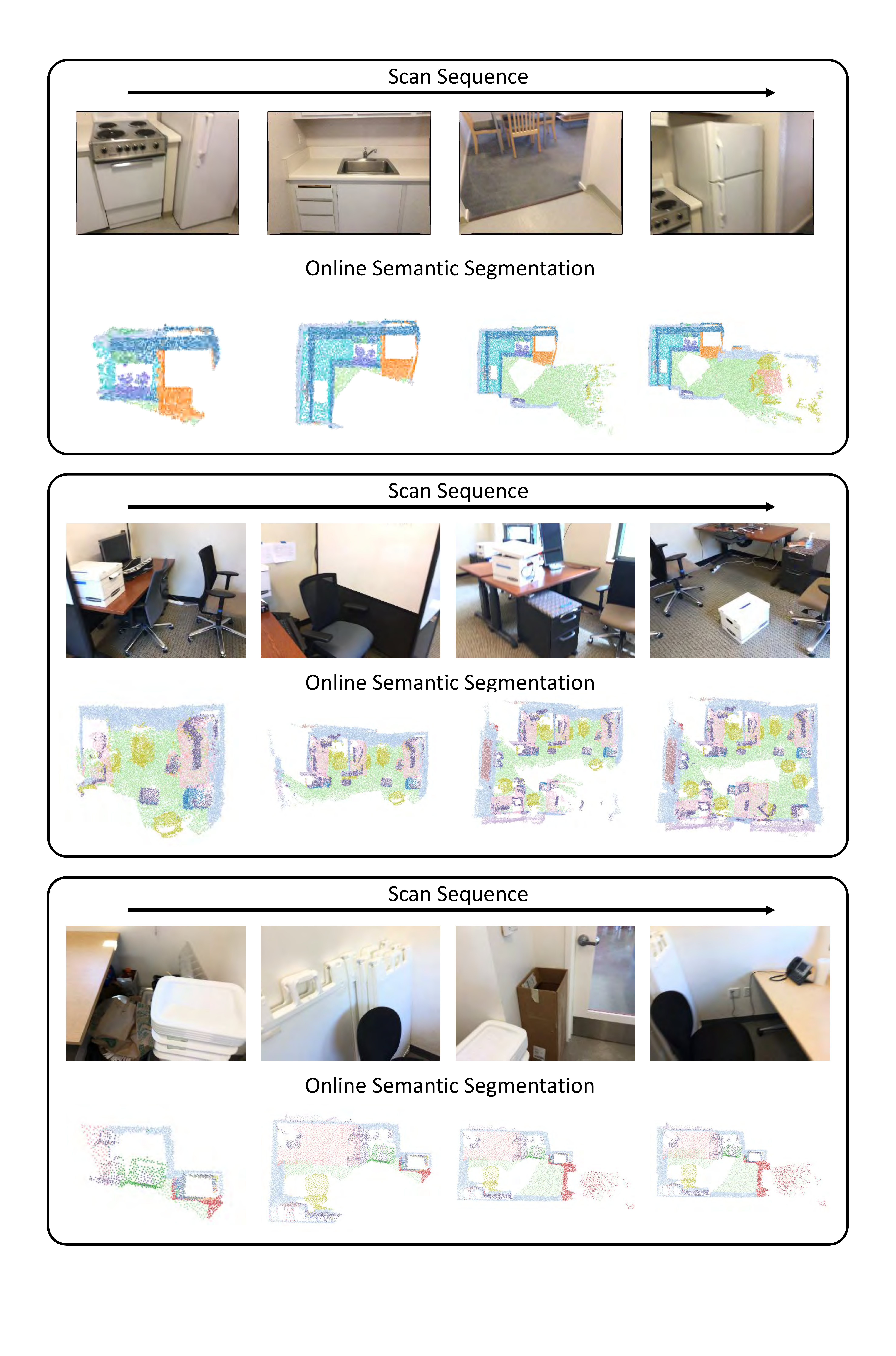}\myfigurename{}
	\end{overpic}
	\caption{
		Visual results of our online semantic segmentation method. Our method can work properly even the input scan is incomplete. 
	}
	\label{fig6}\vspace{-12pt}
\end{figure*}

\begin{figure*}[]
	\centering
	\begin{overpic}
		[width=0.85\linewidth]
		{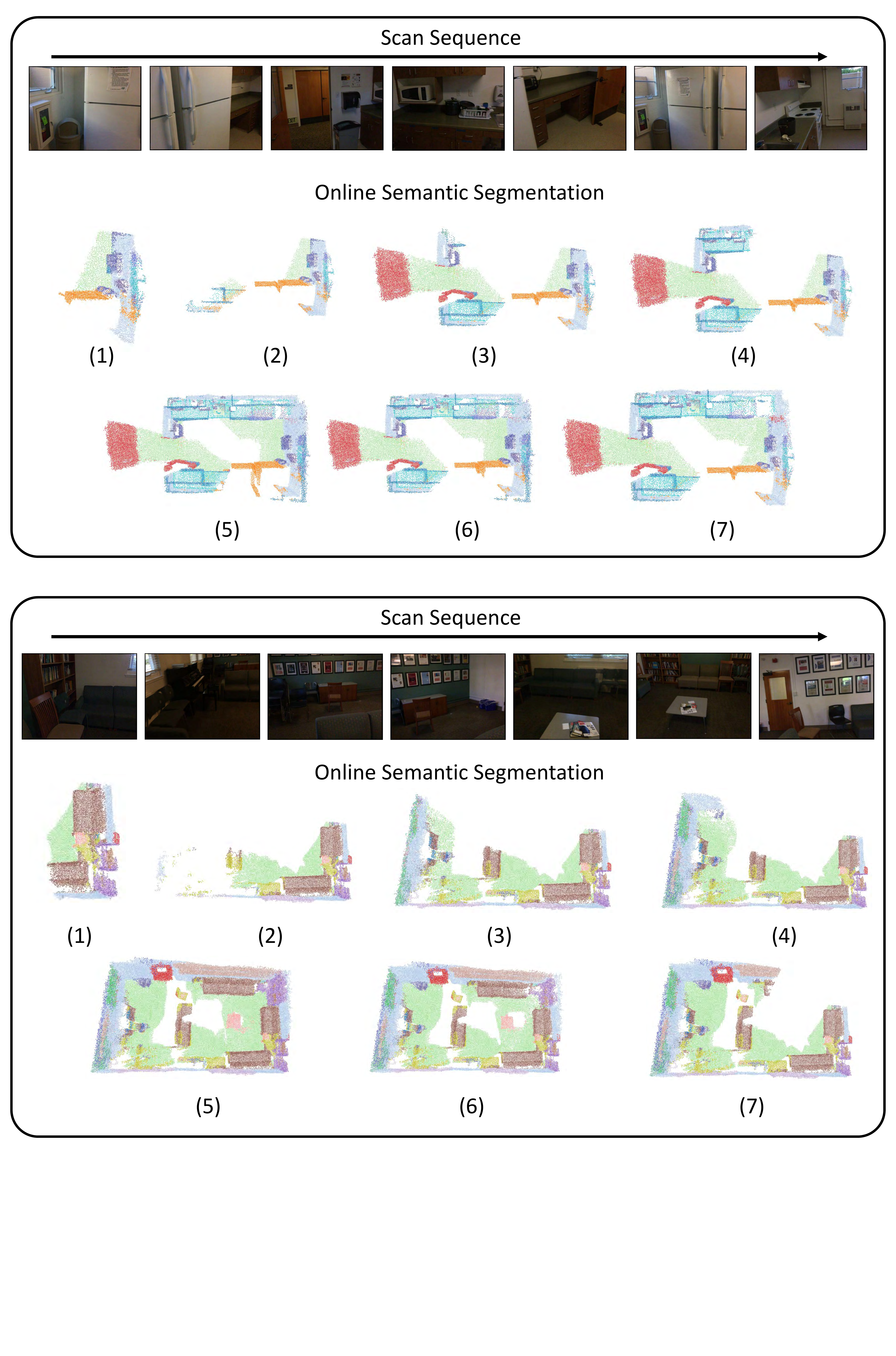}\myfigurename{}
	\end{overpic}
	\caption{
		Visual results of our online semantic segmentation method. Our method can work properly even the input scan is incomplete. 
	}
	\label{fig7}\vspace{-12pt}
\end{figure*}

\begin{figure*}[]
	\centering
	\begin{overpic}
		[width=0.85\linewidth]
		{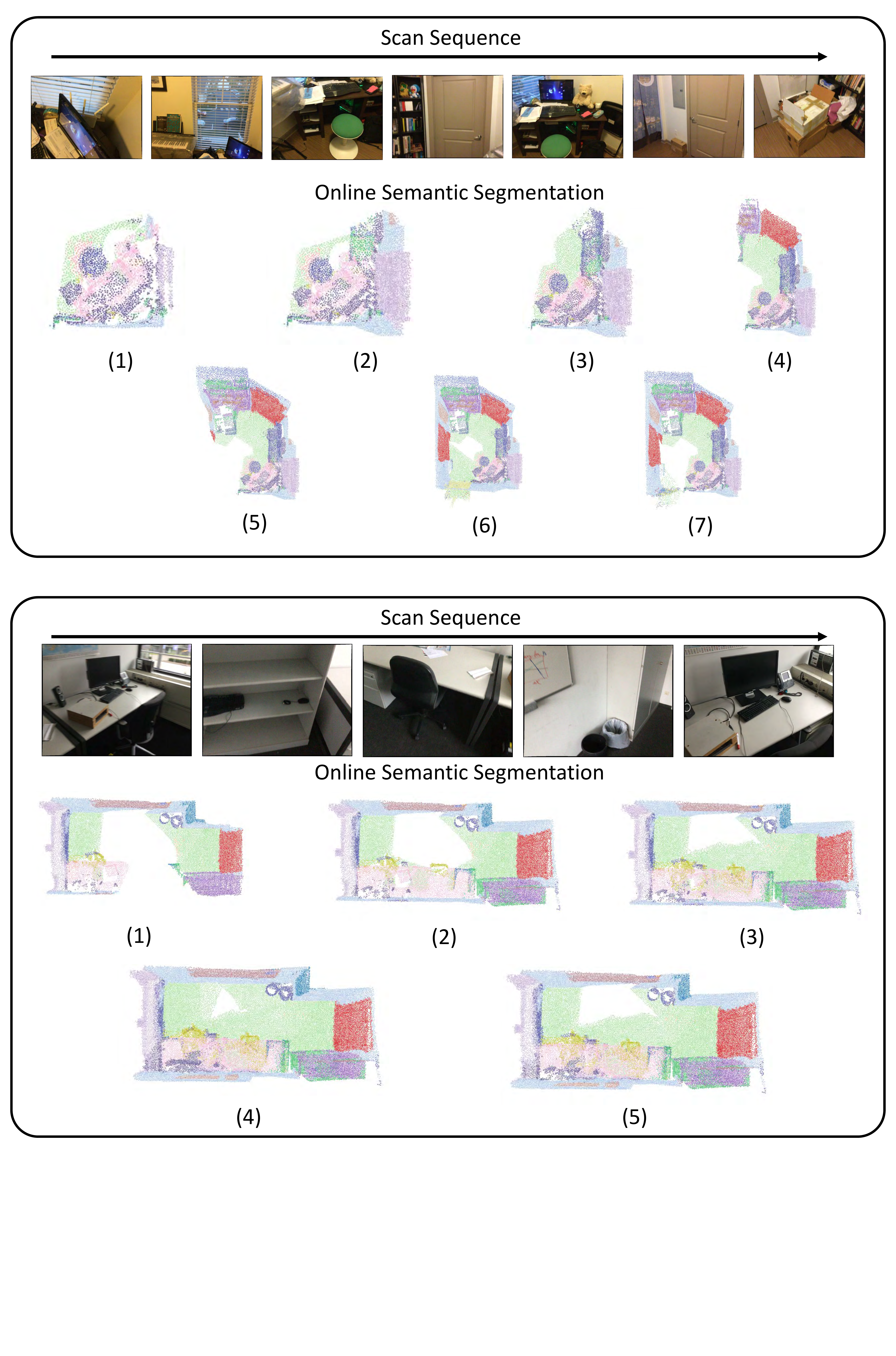}\myfigurename{}
	\end{overpic}
	\caption{
		Visual results of our online semantic segmentation method. Our method can work properly even the input scan is incomplete. 
	}
	\label{fig8}\vspace{-12pt}
\end{figure*}

\begin{figure*}[]
	\centering
	\begin{overpic}
		[width=0.85\linewidth]
		{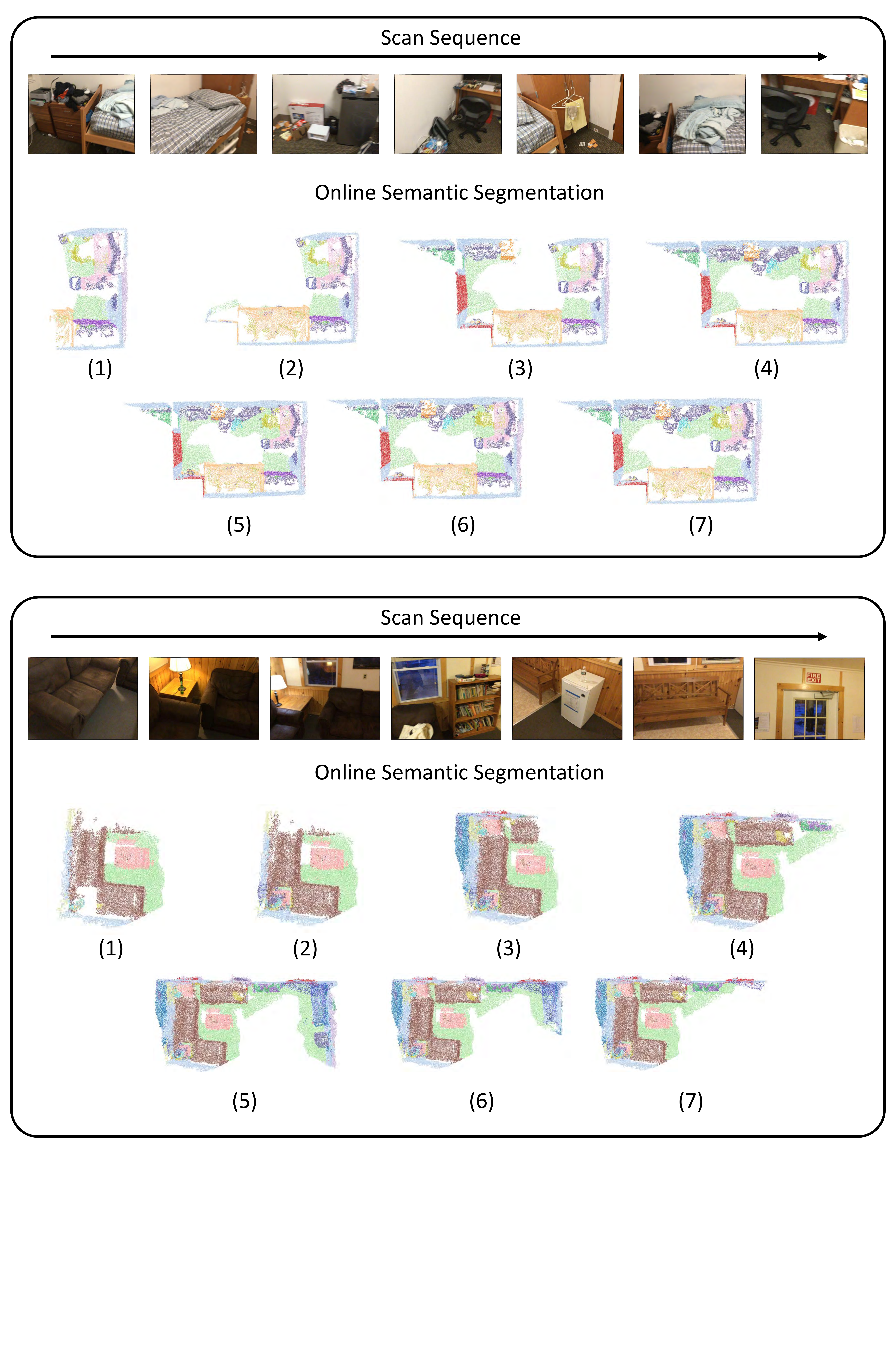}\myfigurename{}
	\end{overpic}
	\caption{
		Visual results of our online semantic segmentation method. Our method can work properly even the input scan is incomplete. 
	}
	\label{fig9}\vspace{-12pt}
\end{figure*}

{\color{black}Figure~\ref{fig:comparsion2d} shows the comparsion of full pipeline 2D model\cite{Hazirbas2016FuseNet} with our methods.}
Figure~\ref{fig1} to \ref{fig9} presents some visual results of our online semantic segmentation method on ScanNet Dataset. Note that, test data shown here selected from the validation and test set in ScanNet Dataset. For live demo, please refer to the attached video in the supplemental material.

\end{document}